# Landau Polarons as Generators of Quantum-Coherent States

Arnab Ghosh[1], Patrick Brosseau[1], Dmitry N. Dirin[2,3], Rui Tao[2,3], Maksym V. Kovalenko[2,3], and Patanjali Kambhampati[1]*

[1]Department of Chemistry, McGill University, Montreal, H3A 0B8, Canada

[2] Department of Chemistry and Applied Biosciences, ETH Zürich, Switzerland

[3] Laboratory for Thin Films and Photovoltaics, Empa - Swiss Federal Laboratories for Materials Science and Technology, Switzerland

*pat.kambhampati@mcgill.ca

**Abstract**

Since Landau's theory, polarons have been understood as quasiparticles in which charges are dressed by the lattice field, yet decades of transport and spectroscopic studies have yielded only static indirect renormalizations. Whether such dressing can dynamically reorganize electronic spectra to generate new quantum-coherent states has remained unresolved. Here we use femtosecond coherent multidimensional spectroscopy on size- and composition-controlled perovskite quantum dots to track polaronic field-induced dynamics in real time, revealing their consequences. We observe a delayed condensation into a confined spectrum of coherent states on 50–150 fs timescales, with couplings between these states evolving dynamically on the same timescale. The splittings are robust, exhibit anomalous linear size dependence, exceed single-particle splittings and manifest at 300 K. A Raman-constrained spin–boson Hamiltonian captures both the anomalous scaling and dynamical onset, establishing polarons as generators of coherent manifolds that enable collective quantum phenomena including superradiance, superfluorescence and superabsorption.

Landau's theory of the polaron, introduced in 1933, remains one of the cornerstones of condensed matter physics [1]. By showing that an electron or hole becomes dressed by lattice vibrations, Landau established that charge carriers are not bare particles but composite quasiparticles whose mass and interactions are renormalized by the surrounding medium. This simple but profound idea inspired decades of theoretical elaboration. Fröhlich extended the model to long-range polar coupling [2], Holstein described local lattice distortions [3], and Feynman provided a variational treatment of the polaron problem [4]. Polarons have since been invoked across physics to explain colossal magnetoresistance in manganites [5], charge localization in organics [6], and even pairing tendencies in unconventional superconductors [7].

The importance of polaron physics has only grown in the modern era. In correlated oxides, polarons are central to understanding localization–mobility competition and emergent phases [8]. In molecular crystals and organic semiconductors, polaron formation shapes charge mobility, exciton transport, and light–matter coupling [9]. Most recently, halide perovskites have emerged as a uniquely revealing system. Their highly polarizable lattices, soft phonon modes, and long carrier lifetimes have led to proposals that polarons underpin defect tolerance [10], hot-carrier bottlenecks [11], and diffusion lengths exceeding microns [12]. Ultrafast structural probes have revealed large-amplitude lattice responses [13,14], while femtosecond optical spectroscopies [15-19], including our own [20-24], have begun to directly capture coherent polaron formation on sub-picosecond timescales. These advances have pushed the field beyond static inference, showing that polaron physics in perovskites unfolds dynamically.

Yet a central element has remained missing. While polaron formation itself has now been observed, its consequences—the new states, couplings, and coherences that emerge as the polaron cloud builds—have not been resolved. Traditional signatures such as linewidth broadening, effective mass enhancement, or phonon sidebands [7,25] describe only the aftermath. What has been lacking is the ability to watch the spectrum reorganize in real time: do polarons merely renormalize existing states, or can they generate entirely new manifolds? This question echoes other discovery moments in condensed matter physics where new tools revealed that interactions create states of matter rather than perturb them. Just as femtosecond lasers exposed coherent phonon dynamics [26] and ultrafast electron diffraction revealed nonthermal phases of solids [26], so too must new ultrafast spectroscopies uncover what polarons actually do.

Here we show that exciton–polaron interactions in halide perovskite nanocrystals generate a dynamically confined manifold of excitonic states—quantum drops—that emerge as the polaron cloud forms. Using femtosecond coherent multidimensional spectroscopy, we capture delayed spectral formation with dynamically evolving energies and couplings. The splittings are large, persist to 300 K, and exceed the observed single-particle splittings versus size. The effects are robust in both $CsPbBr_3$ and $CsPbI_3$, with the magnitude reflecting the strength of exciton–polaron coupling. These states overturn the textbook $1/D^2$ scaling of confinement, instead displaying linear size dependence and femtosecond onset on 150–300 fs timescales. A microscopic spin–boson Hamiltonian constrained by Raman spectra quantitatively reproduces the scaling and dynamical onset, confirming that the confining potential is generated by the polaron itself.

Taken together, these results represent the first direct observation of the consequences of polaron formation. Polarons are not passive dressings but active generators of many-body physics, producing quantum-coherent, dynamically evolving manifolds of states and couplings. Moreover, the correlated lattice fluctuations that create polaron confinement also stabilize long-lived coherence, linking lattice dynamics to persistent quantum order. This establishes polaron-induced dynamical confinement as a new organizing principle for quantum materials, extending Landau's legacy into a regime where interactions create confinement, coherence, and collective states unavailable to static lattices.

The concept of polarons as a platform for localizing excitons is presented in **Fig1. Fig1a** shows a schematic illustration of a crystal lattice prior to excitation on the left, and after excitation and relaxation on the right. For a polar lattice at equilibrium, there will be partial charges illustrated by the red/blue ellipses, aligned in some ordered manner. The lattice may undergo motion in some direction as shown by the vertical arrows corresponding to a phonon displacement field. For soft, anharmonic lattices, the introduction of a central charge creates lattice displacements about this centroid. The lattice motions do not follow a single normal mode, but instead follow a local mode for a displacement field of high divergence. The presence of an exciton is known to give rise to couplings to polarons[9,13,15,17,20,21,27-29]. But is the reverse possible, in which coupling to polarons gives rise to quantum confined excitons?

A candidate for observing new physics of polaron induced quantum localization of excitons is the lead halide perovskite (LHP) quantum dot (QD) [30-32], due to their unique lattice in which a dynamically disordered liquid-like lattice, is strongly coupled to charges. **Fig1b** shows a transmission electron microscope (TEM) image of 18 nm edge length $CsPbBr_3$ LHP QD. The polaron diameter and the exciton Bohr diameter are both ~ 7 nm, illustrated by a yellow sphere in the center of the QD image. It is clear from the size scales that a large 18 nm LHP QD is not a standard physically confined QD. Yet these large QD still support some QD-like physics such as single photon emission[33,34]. Since the exciton Bohr diameter and the polaron diameter are nearly the same, one anticipates the possibility of exciton-polaron coupling giving rise to emergent quantum systems. **Fig1c** shows the lattice structure of these LHP QD, which exhibits an orthorhombic crystal structure at room temperature.

The electronic density of states is reflected in the linear absorption spectra, shown in **Fig1d** for these 18 nm PQD. In this bulk QD size regime there is a band edge exciton observed at 300K, but no evidence of an excitonic manifold. **Fig1e** shows the differences between a phonon and a polaron limit of lattice dynamics. Shown are the Raman power spectra for CsPbBr3 QD obtained by coherent multi-dimensional spectroscopy (CMDS), Raman, and ab initio molecular dynamics (AIMD) [24,34]. Also shown is the power spectrum of 3.9 nm diameter CdSe QD, obtained via transient absorption (TA) spectroscopy. The LHP QD Raman spectra are consistent with each other, in revealing a low frequency spectral density. This spectral density arises from the diffusive lattice dynamics characteristic of polarons. These dynamics are particularly notable for being liquid-like, representing polaron formation [20,22,35]. In contrast, the CdSe QD Raman spectrum is comprised of two well-defined Lorentzian peaks. There is an acoustic and an optical mode. Here, the lattice dynamics moves in a coherent, underdamped manner. These Raman spectra support the proposal of polaronically induced localization of excitons.

**Fig2** provides a detailed overview of how polaron formation might give rise to spectroscopic observables that unambiguously demonstrate quantum size effects having a turn-on time. **Fig2a** shows a configuration coordinate diagram of a polaronic well at a snapshot in time. Represented here is the idea of exciton-polaron coupling [15,17,20,27-29,35]. Excitation from the ground state (not shown) into the exciton state results in polaron formation, with a coupling strength derived from the relative displacement of the potential energy surfaces. **Fig2b** illustrates the exciton-polaron system after completion of polaron formation. The timescale of this process is denoted by the first time step, $\Delta t_1$. Initially there was one excitonic state, in which the bulk exciton energy is dressed by this polaronic zero-point energy. With a deeper polaronic well, a second or more excitonic states can form at higher energies, denoted by the level in blue. The former lowest exciton state experiences additional blueshifting by polaronic confinement, moving the red level to orange. These processes are all dynamical, expected to evolve on the 150 fs timescale of polaron formation in LHP [9,20-22,27-29,35].

Direct evidence for polaron induced formation of quantum confined excitons comes from the observation of dynamic blueshifting of the lower exciton, $X_1$, in concert with the delayed appearance of an upper exciton, $X_2$. A quantum confined excitonic system is not just characterized by a spectrum of states, but couplings between these states [36]. **Fig2c** illustrates the idea that the two exciton states are coupled, at some delayed time. The timescale of this process is denoted by the second time step, $\Delta t_2$.

**Fig2d-f** schematically illustrates a spectroscopic measurement that can resolve excitonic spectra and couplings, in energy and time. Shown is a two-dimensional spectrum with an excitation and a detection axis. At early time there is a single peak along the diagonal at low energy, labeled in red. At late time there are two peaks along the diagonal now. The red lower peak is now blueshifted to orange. And the second peak appears in blue. These two exciton states form an uncoupled spectrum. In the case of couplings between these exciton states, there will be peaks along the anti-diagonal as shown in green.

This two-dimensional correlation map is obtained by coherent multi-dimensional spectroscopy (CMDS) [37,38], schematically illustrated in **Fig2g.** The method involves four-wave mixing of three phase coherent incident fields. The linear polarization generated by the first field, is correlated to the third order nonlinear polarization created by the third field. The first field, $E_1$, generates the excitation axis, and the third field, $E_3$, generates the detection axis. The second field transforms the system from a coherence to a population, which undergoes dynamics to be later correlated by the third field. The radiated polarization is heterodyne detected and spectrally resolved, as discussed in the Supplemental Information (**SI**), and our prior CMDS work [20,22-24,35,39,40].

The theoretical framework for the observed dynamics is developed from an effective spin–boson Hamiltonian [41], explicitly incorporating the coupling of excitons to a Raman-active phonon bath, as described in detail in the **SI**. In this approach, the excitonic states are represented as a two-level system coupled to a continuum of bosonic modes that capture the phonon environment. The system Hamiltonian governs the excitonic energies and splittings, while the bath Hamiltonian describes the lattice vibrations, and the interaction term parameterizes the exciton–polaron coupling. The phonon bath is constrained by the experimentally measured Raman spectral density, thereby ensuring that the model is not phenomenological but grounded in directly observed vibrational structure. Equations of motion for the reduced density matrix are then derived, incorporating the coherent buildup, size dependence, and subsequent dephasing of the splittings.

**Fig3** shows CMDS data on 18 nm edge length $CsPbBr_3$ bulk QD and 3.9 nm diameter CdSe strongly confined QD. Spectra are shown at three population times, $t_2$ = 0, 100, 1000 fs. These population times were chosen to capture glimpse of the exciton-polaron system prior to, during, and following completion of polaronic well formation. The CMD spectra for $CsPbBr_3$ bulk QD show a single peak at time zero, corresponding to an exciton coupled to a polaron in its first stages of birth, **Fig3a**.

During polaron formation on the 150 – 200 fs timescale, CMD spectra are shown. Now there is a clear second peak with a delayed formation. On this timescale, there is a clear transformation from the conventional idea of an exciton-polaron, to a polaron induced quantum dot. Following completion of polaron formation at $t_2$ = 1000 fs, the CMDS spectrum is further transformed from having two peaks to four. The peaks along the energy diagonal (D) correspond to a density of states for an excitation spectrum. The peaks along the anti-diagonal (AD) correspond to the couplings between excitonic states along the diagonal.

In the first stages of polaron formation, the exciton–polaron complex evolves into a polaron-induced quantum dot, in which the exciton spectrum is reshaped by the emergent lattice potential. This regime is characterized by dynamic confinement, where the effective well depth increases with dot diameter and composition-dependent coupling strength. In the final stages of polaron formation, however, the system undergoes a further transformation: the polaron-induced quantum dot evolves into a coupled excitonic phase, a distinct class of quantum dot where the exciton is no longer confined by the static lattice boundary but by the dynamically generated polaron cloud. Each of these distinct quantum phases—self-confined exciton–polarons and their eventual transition into a collective coupled excitonic state—is explicitly predicted by theory using the effective spin–boson Hamiltonian constrained by Raman spectral densities. The equations of motion thereby capture the entire progression of the dynamics, from initial self-trapping to the establishment of a new excitonic phase.

These LHP bulk QD reveal that in a lattice that supports liquid-like structural dynamics, polaron formation gives rise to a new form of QD. **Fig3b** shows the behavior of the canonical CdSe strongly confined QD. Here, there are four peaks observed at all population times. CMDS reveals that CdSe in a 3.9 nm diameter nanocrystal indeed forms a static, physically confined QD with no dynamics beyond simple hot exciton cooling [39].

The first prediction from **Fig2** and detailed by theory is the delayed onset of a spectrum of quantum confined excitons. **Fig3c** illustrates the dynamics of this polaron induced QD formation by monitoring the fraction of $X_2$ over time. Shown is the normalized amplitude of the area under $X_2$ relative to the total of $X_1$ and $X_2$, the total signal. The LHP QD clearly shows a delayed onset to the appearance of $X_2$ on the 150 fs timescale of polaron formation. This step, $\Delta t_1$ is complete in 80 fs. The CdSe QD is static as expected for a conventional physically confined QD. This delated onset of a higher-lying exciton is clear evidence of polaron induced formation of a quantized exciton spectrum that is characteristic of a strongly spatially confined QD.

The second prediction from **Fig2** and theory is that the energy of the lowest exciton, $X_1$, will show a time-dependent blueshift due to strong exciton-polaron coupling. **Fig3d** shows the peak shift of the $X_1$ state over time. Peak shifts were obtained for an excitation window on the red edge of the band edge exciton to minimize broadening effects [35]. For the LHP QD there is a clear blueshifting of 5 meV on the same 150 fs timescale of polaron formation. This observation is further corroboration of the emerging picture of strong-exciton-polaron interactions giving rise to new quantum effects. In contrast to the LHP QD, the CdSe QD shows a prompt redshift of 5 meV, arising from relaxation in the exciton fine structure [39]. The theory explains the 5 meV blueshift on the 150 fs timescale as the direct consequence of rapid polaron formation. In the first stage of the dynamics, the exciton strongly couples to the polar lattice, and the coherent reorganization of the lattice potential renormalizes the exciton energy upward, producing a transient blueshift. The effective spin–boson Hamiltonian with a Raman-constrained spectral density predicts this response quantitatively, because the bath modes that dominate on ~100–200 fs timescales supply the necessary lattice energy to dynamically deepen the confinement potential, shifting the exciton resonance by ~5 meV.

The third prediction from **Fig2** is the possibility of the delayed formation of coupling between the basis exciton states in the new manifold. **Fig4** illustrates how couplings are observed in CMDS. As shown in **Fig4a**, a two-dimensional spectrum can be characterized by the splittings between the diagonal peaks (D) and the anti-diagonal (AD) peaks. The AD peaks are discussed in the CMDS literature as revealing couplings between states [38] [37]. **Fig4b** shows the AD spectrum over time. At time zero, the AD spectrum shows a single peak at the center of the spectrum, revealing the complete absence of coupling because the upper state has not yet formed. Only after 100 fs, the onset of exciton spectrum formation, does there appear to be a true AD spectrum. The AD spectrum is narrow initially and then broadens over the following 200 fs. By 300 fs the AD spectrum reveals two peaks, corresponding to the couplings between exciton states now being fully formed.

The dynamics of these splittings are shown in **Fig4c**. The D splittings are constant, indicating that the blueshifting of $X_1$ also causes synchronous blueshifting of $X_2$, with the transition energy being conserved. In contrast, the AD splittings show a clear delayed formation followed by a dynamical splitting that reveals the dynamics of exciton coupling during formation of this composite excitation. The arrows show the first step of spectrum formation in the D splittings and now the second step of coupling formation. This step, $\Delta t_2$ is complete in 220 fs. The same AD spectral analysis was performed on CdSe QD (**SI**). CdSe QD reveals none of this dynamical coupling behavior. In CdSe QD, all the couplings and splittings are static, consistent with a static physically confined QD. The theory explains the anti-diagonal (AD) dynamics as the gradual buildup of coherent exciton–polaron correlations. At early times (<80 fs), no AD signal appears because the polaron cloud has not yet formed and the exciton spectrum remains uncorrelated across excitation and detection energies. By ~80 fs, the partially developed polaron bath produces a transient central AD peak, reflecting nascent coherent coupling. By ~300 fs, full polaron formation splits this central feature into two distinct AD peaks, corresponding to the stabilized exciton–polaron eigenstates predicted by the effective spin–boson Hamiltonian with Raman-constrained dynamics.

While bulk LHP QD of 18 nm edge length were employed to show that these dynamical quantum confinement processes, a complete size dependence can further support these effects being non-trivial, arising from new effects rather than conventional size effects. The size dependence of these splittings is shown in **Fig5** and compared to other size dependent phenomena.

**Fig5a-b** show the D and AD splittings vs edge length. Data is shown for a wide size range of $CsPbBr_3$ QD, synthesized by Kovalenko [33,42]. Data is also shown for 15 nm $CsPbBr_3$ and $CsPbI_3$ QD, synthesized by Kambhampati [35]. The size dependence follows a linear functional form for both D and AD splittings. The D splittings increase from 40 to 80 meV as the QD increases from strongly confined to bulk. The AD splittings increase from 25 to 60 meV. Remarkably the data from 15 nm QD synthesized by a different group and measured at a different time[35], perfectly fits on the linear curve from these new data. The $CsPbI_3$ data reveals considerably larger splittings, consistent with the difference in polaron formation time. [35].

These anomalous splittings in the nonlinear spectroscopy of LHP QD are compared to the conventional exciton splittings observed in the linear spectroscopy of LHP QD in **Fig5c**. Shown are the splittings for the band edge exciton energy (1S), above the bulk value. Also shown are the SP splittings at 300K and at 15K [43]. As expected for any physically confined QD, the splittings follow an inverse size dependence, to some power. Notably, over this size range the linear QD size effects based splittings are from 10 to 100 meV. With this reference energy scale, the polaron induced splitting is as large or larger than the physical confinement-based splitting. And the dynamic blueshifting is as large as the energy of the band edge exciton above the bulk value. These polaron induced splittings are large, relative to standard QD size effects based splittings.

There have been other observations of anomalous size dependence to spectroscopic observables in LHP QD. **Fig5d** shows the photoluminescence (PL) rate constant vs. edge length for $CsPbBr_3$ QD. The data is adapted from size-dependent single QD PL measurements by Kovalenko[33] and ensemble PL on one size by Kambhampati[44]. These experiments reveal that a spatially coherent state forms at low temperature, enabling single photon superradiance. The main point is that these effects arise in bulk QD by virtue of some new confinement mechanism.

This new dynamic confinement mechanism is illustrated in **Fig5e**. Drawn to scale are LHP QD of 20, 10, 5 nm, corresponding to three different confinement regimes by this polaronic potential. The polaron is indicated by a 7 nm diameter sphere, illustrating the lattice distortion field. Within each polaron cloud is the wavepacket that emerges from a coherent superposition of these two exciton states[24]. The larger LHP QD has larger splittings, giving rise to a long-lived coherent wavepacket[24] rendered in blue to illustrate the large splitting. As the QD becomes smaller, the polaronic confinement becomes weaker, resulting in lower energy wavepackets in green and red. Notably the amplitude of these long-lived excitonic wavepackets is also linearly size dependent[24].

The theory predicts that the diagonal (D) splittings increase linearly with size because the confinement potential is not set by the static nanocrystal boundary but instead emerges from the dynamically growing exciton–polaron cloud, whose collective interaction strength scales with exciton coherence volume. The composition dependence arises because $CsPbI_3$, with its softer lattice and stronger polarizability, supports more robust exciton–polaron coupling than $CsPbBr_3$, leading to larger splittings. The anti-diagonal (AD) splittings are smaller because they reflect only the dynamic coherence transfer mediated by the Raman bath, which produces weaker off-diagonal correlations compared to the static diagonal self-energy shifts. The microscopic parameters obtained from theory are listed in the **SI.**

Since Landau first introduced the concept of the polaron [1], it has stood as a fundamental paradigm for how carriers are renormalized by their environment. What had remained elusive, however, were the direct consequences of polaron formation. Here we have resolved them for the first time. By capturing the delayed spectral birth of quantum drops—polaron-confined excitonic states—we show that polarons do not simply dress excitons but actively generate new, coherent many-body manifolds of states and couplings. These states are robust, large in magnitude, persist at 300 K, and exceed the observed single-particle splittings versus size, demonstrating that polaronic confinement is both a universal and technologically relevant effect.

Beyond confinement, the coherence stabilized by correlated lattice fluctuations connects this work to a broader frontier of quantum optics in solids. The coherent many-body states we observe resonate with collective effects already reported in perovskite nanocrystals—superradiance [33,44], superfluorescence [45], and superabsorption [43] —and establish a microscopic mechanism by which such coherence can be stabilized and amplified. In this way, polaron physics links directly to phenomena central to quantum photonics, entanglement, and quantum information science. By extending Landau's original vision into the regime where polarons generate—not just renormalize—states of matter, we establish polaron-induced dynamical confinement as a new organizing principle for quantum materials, and provide a foundation for engineering perovskites as a platform for coherent quantum technologies.

**Online content:**

Any methods, additional references, Nature Portfolio reporting summaries, source data, extended data, supplementary information, acknowledgements, peer review information; details of author contributions and competing interests; and statements of data and code availability are available at:

**Figures.**

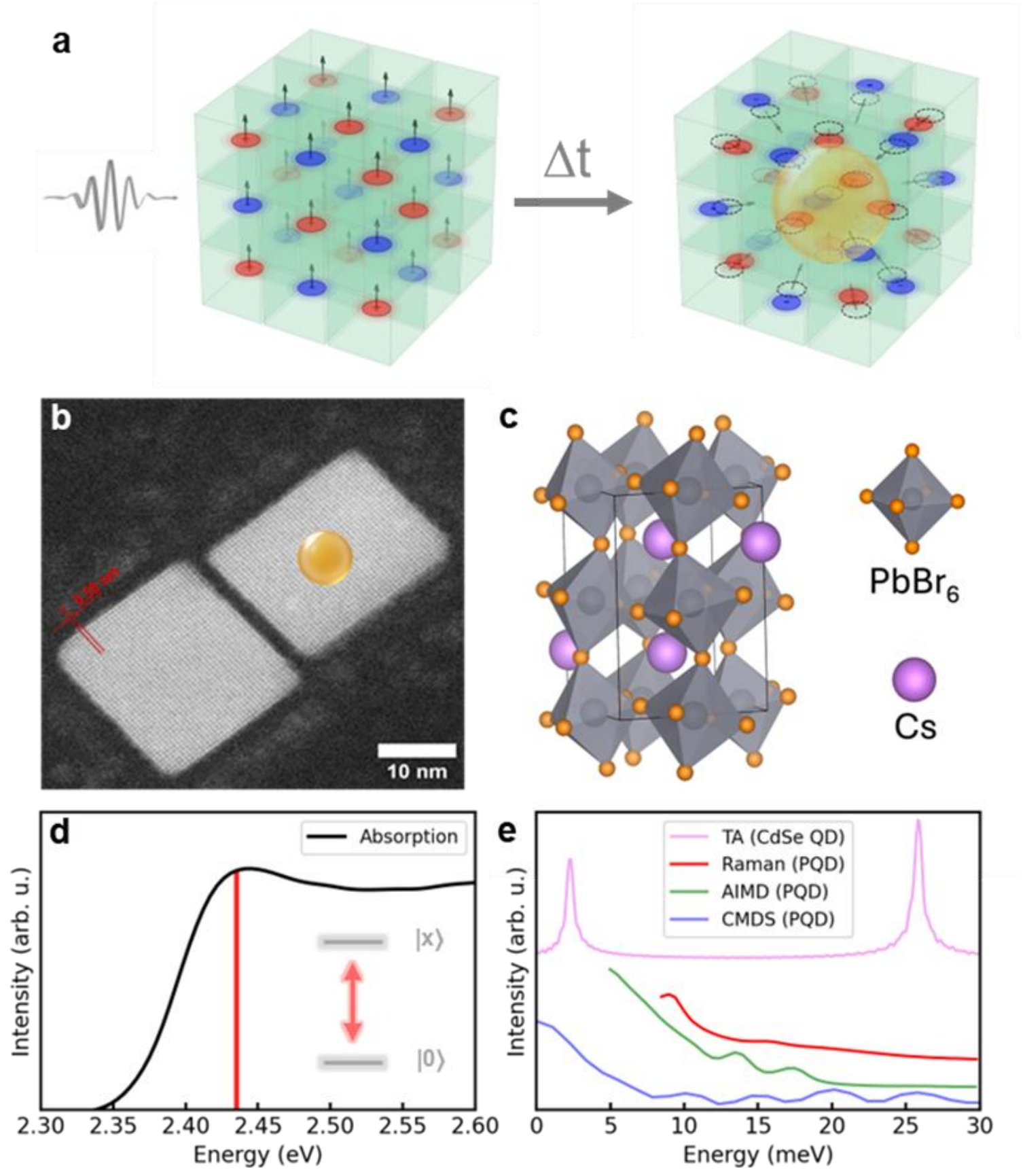

**Fig1. Introduction to the concepts of polaron induced charge localization.** a) Schematic illustration of photon induced lattice dynamics giving rise to a quantum confining potential in the absence of geometrical constrains. b) Transmission electron microscope (TEM) image of an 18 nm edge length bulk $CsPbBr_3$ perovskite quantum dot (PQD). The yellow sphere represents the polaron distortion field. c) Illustration of the lattice structure of this perovskite. d) Linear absorption spectrum of the 18 nm PQD. Raman spectra of $CsPbBr_3$ QD and 3.9 nm diameter CdSe QD. Data are adapted from [24,34] with permissions.

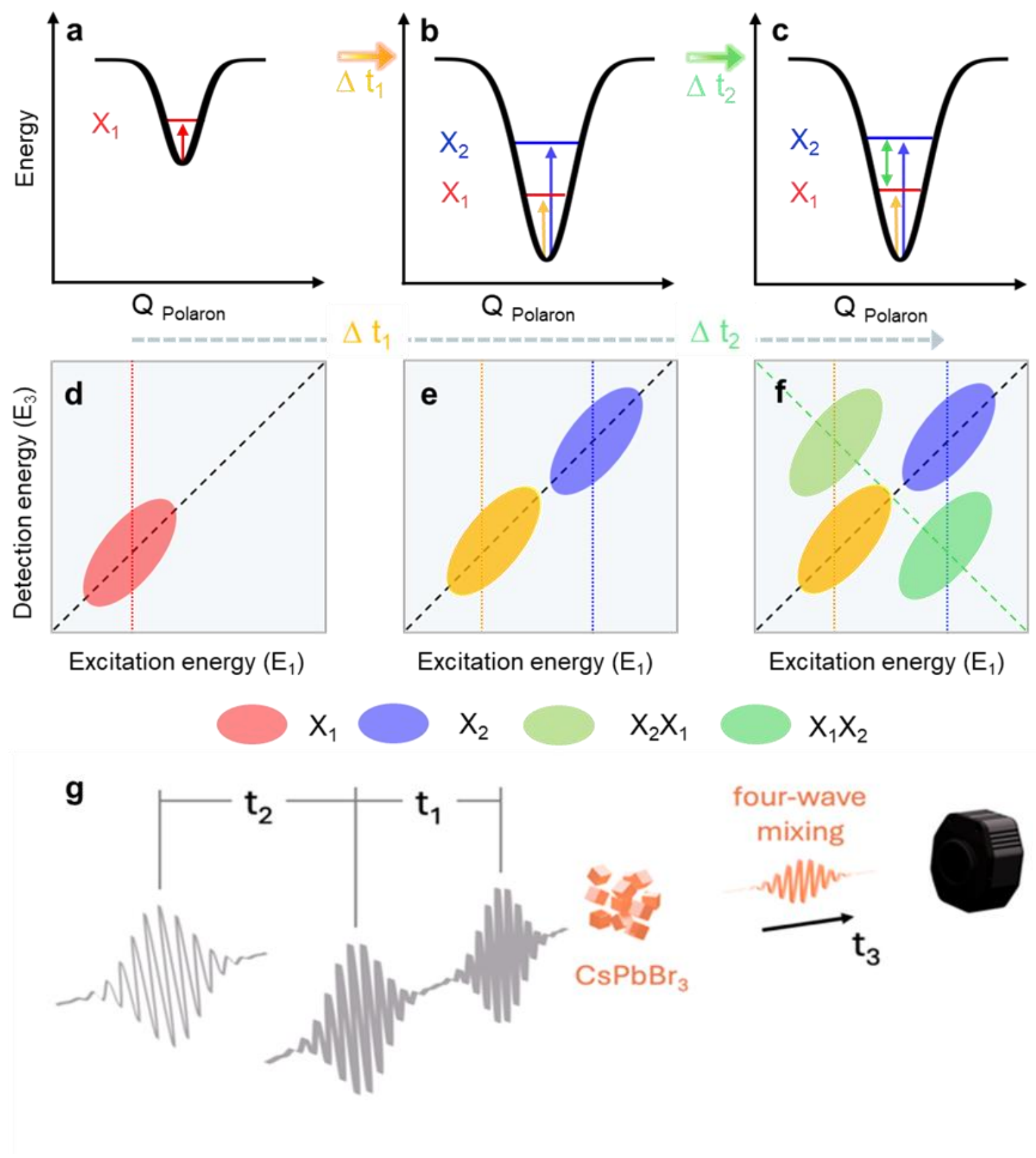

**Fig2. Illustration of how polaronic charge localization into quantized states may be experimentally observed.** a) Initial formation of an exciton-polaron. b) A time-delayed increase the depth of the polaronic potential may give rise to a spectrum of quantum confined excitonic states. c) Upon further time evolution, the excitonic states may consist of coupled excitons. d) A two-dimensional map of excitation and detection energy will have a characteristic peak for an exciton-polaron. e) The same 2D map for a polaronically confined exciton spectrum. f) The 2D map, with illustration of couplings between states. g) These 2D energy maps are experimentally measured using Coherent Multi-Dimensional Electronic Spectroscopy (CMDS).

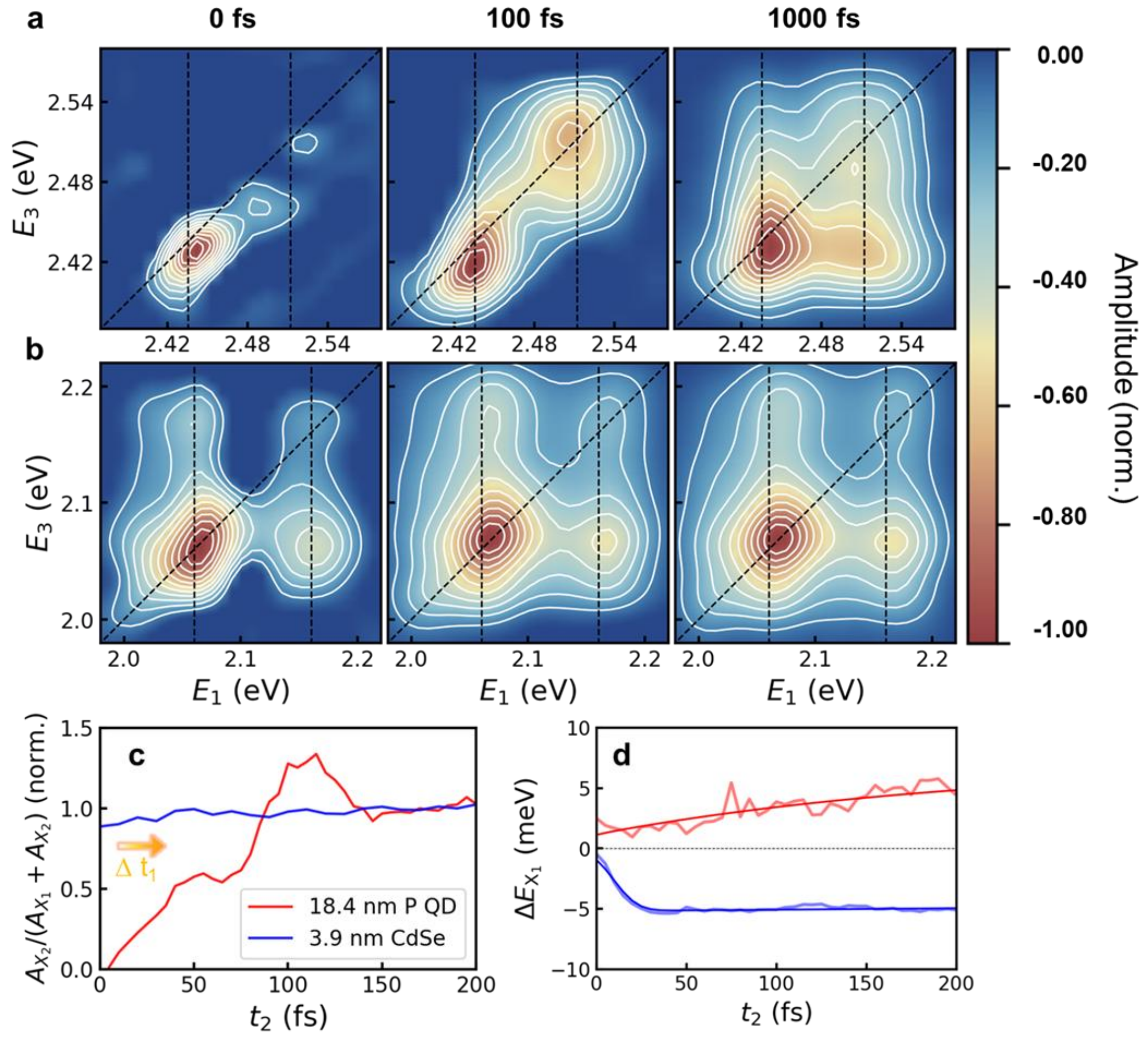

**Fig3. CMDS reveals the formation of a polaronically induced excitonic spectrum.** a) Absorptive CMDS data of 18 nm edge length $CsPbBr_3$ QD, at three population times ($t_2$). b) CMDS data of 3.9 nm diameter CdSe QD, at three population times ($t_2$). c) The tine dependent fraction of population in $X_2$ as a function of the total population, normalized at long time. The orange arrow notes the formation time of the quantum confined excitonic spectrum, $\Delta t_{1.}$ d) The $X_1$ peak shift dynamics.

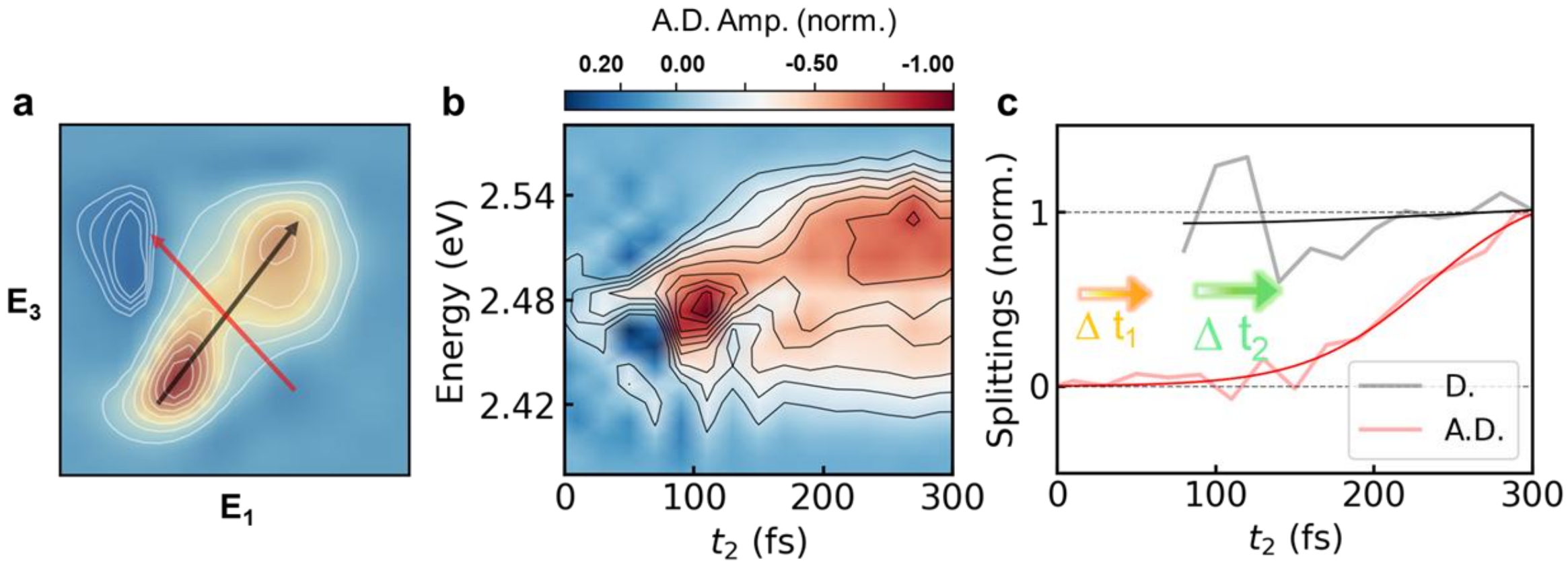

**Fig4. CMDS reveals that couplings in $CsPbBr_3$ QD are dynamic.** a) CMDS spectrum with an anti-diagonal (AD) and diagonal (D) projection shown by arrows. b) AD spectral dynamics showing time-dependent couplings that are absent in physically confined QD. c) The AD and D normalized peak splittings vs time. The orange arrow notes the formation time of the quantum confined excitonic spectrum, $\Delta t_1$, and the green arrow notes the formation time of the fully coupled excitonic manifold.

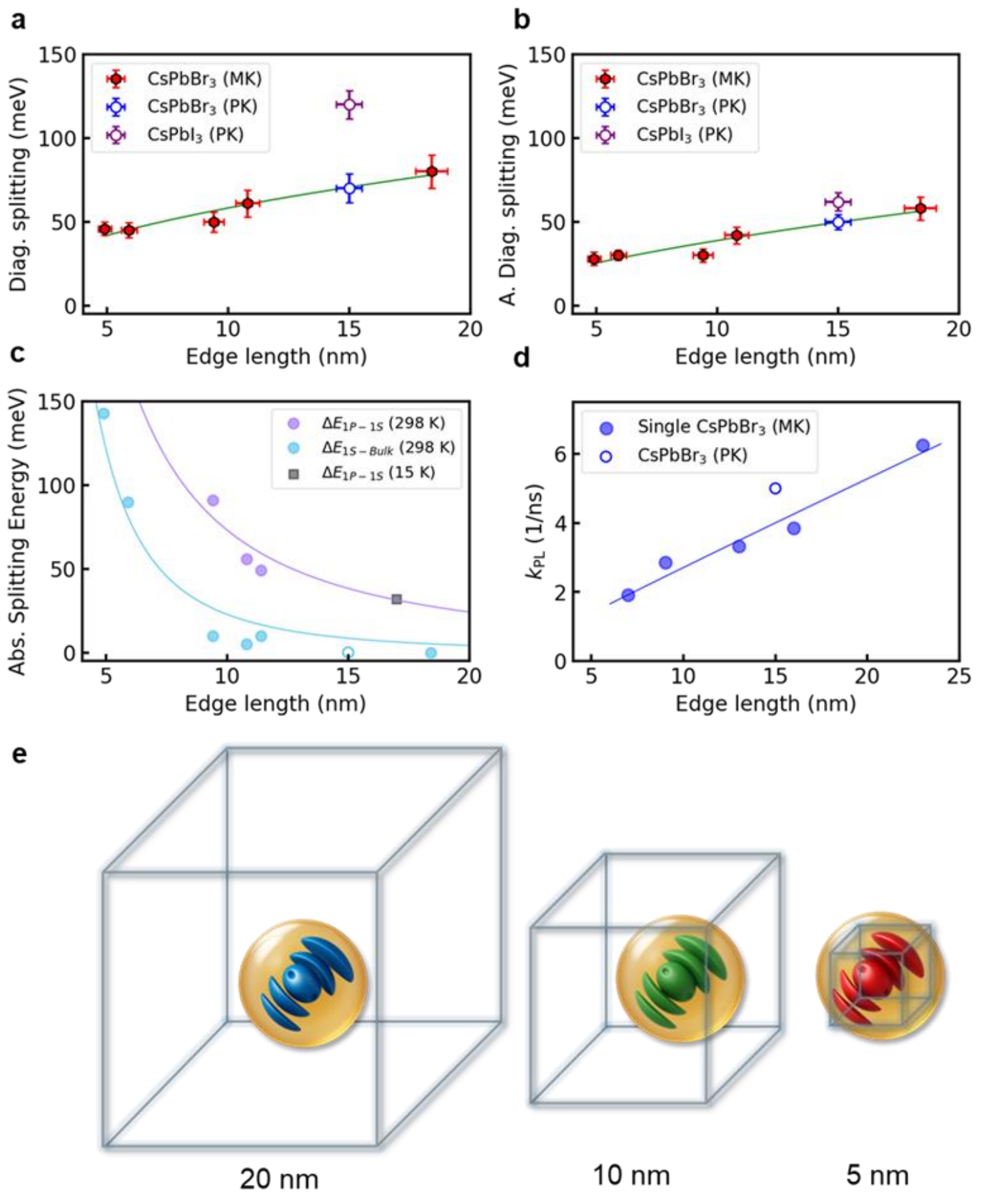

**Fig5. Size-dependent splittings confirm that polaronic localization of charges into quantum confined excitons is a bulk effect.** a) The diagonal (D) peak splittings are size dependent, with the opposite of quantum size effects. b) The anti-diagonal (AD) peak splittings are size dependent, with the opposite of quantum size effects. c) The energy of the band edge exciton above the bulk value, and the SP energy gap, both show the standard size dependence for quantum size effects. The 15K data is adapted from [43] with permissions. d) The radiative rate constant at 4K shows the same anomalous size dependence, due in this case to bulk single-photon superradiance. Data adapted from [33,44] with permissions. e) Illustration of polaron induced formation of quantum dots without boundaries. As the QD becomes larger, there is a greater effect of the polaronic potential. The excitonic effect of this polaronic potential is rendered by the color of the coherent exciton wavepacket that is long-lived [24]. The qualitative picture is exactly modeled by theory.

## References.

**Supplementary Information:**

# Landau Polarons as Generators of Quantum-Coherent States

Arnab Ghosh[1], Patrick Brosseau[1], Dmitry N. Dirin[2,3], Rui Tao[2,3], Maksym V. Kovalenko[2,3], and Patanjali Kambhampati[1]*

[1]Department of Chemistry, McGill University, Montreal, H3A 0B8, Canada

[2] Department of Chemistry and Applied Biosciences, ETH Zürich, Switzerland

[3] Laboratory for Thin Films and Photovoltaics, Empa - Swiss Federal Laboratories for Materials Science and Technology, Switzerland

*pat.kambhampati@mcgill.ca

# Table of Contents

## 1. Synthesis and characterization of lead halide perovskite quantum dots

### 1.1 Synthetic Methods

The lead halide perovskite quantum dots ( LHP QDs) are synthesized using previously published methods [1,2].

### Materials

Lead bromide ($PbBr_2$, 99.999%), cesium carbonate ($Cs_2CO_3$, 99.9%), hexane (≥99%), diisooctylphosphinic acid (DOPA, 90%), oleic acid (OA, 90%) and acetone (ACE, ≥99.5%) were purchased from Sigma Aldrich. n-Octane (min. 99%) and lecithin (>97% from soy) were purchased from Carl Roth. Trioctylphosphine oxide (TOPO, min. 90%) was purchased from Strem Chemicals.

### $CsPbBr_3$ QDs synthesis procedures

0.04 M Pb stock solution. The $PbBr_2$ stock solution (0.04 M) was prepared by dissolving 2 mmol (0.734 g) of lead bromide and 10 mmol (3.866 g) of trioctylphosphine oxide into 10 ml of octane at 120 °C in a 40 ml vial. Once all the $PbBr_2$ was dissolved (~30 min), the solution was cooled to room temperature and transferred to the big Schott bottle and diluted by 40 ml of HEX. The stock solution was filtered after preparation over a 0.45 ul PTFE filter and stored in air.

0.02 M Cs-DOPA stock solution. The Cs-DOPA stock solution (0.02 M) was prepared by loading 100 mg of $Cs_2CO_3$ (0.614 mmol Cs) together with 1 ml of DOPA (3.154 mmol) and 2 ml of OCT at 120 °C in a 40 ml vial. Once all the $Cs_2CO_3$ was dissolved (~20 min), the stock solution was cooled to room temperature and 27 ml of HEX was added. The stock solution was filtered after preparation over a 0.2 ul PTFE filter and stored in air.

0.2 M Cs-DOPA stock solution. The concentrated Cs-DOPA stock solution (0.2 M) was prepared by loading 100 mg of $Cs_2CO_3$ (0.614 mmol Cs) together with 1 ml of DOPA (3.154 mmol) and 2 ml of OCT at 120 °C in a 40 ml vial. Once all the $Cs_2CO_3$ was dissolved (~20 min), the stock solution was cooled to room temperature. The stock solution was filtered after preparation over a 0.2 ul PTFE filter and stored in air.

The lecithin stock solution (~0.13 M) was prepared by dissolving 1 gram of lecithin in 20 ml of HEX using an ultrasonic bath (30 min). After preparation, the stock solution was centrifuged, filtered over a 0.2 ul PTFE filter and stored in air.

Synthesis of $CsPbBr_3$ QDs was performed following the procedure reported elsewhere [2] with slight modifications.

**5.9 nm LHP QDs.** The filtered hexane (48 ml) is mixed with $PbBr_2$ stock solution (8 ml). Under heavy stirring, a 0.02 M Cs-DOPA stock solution was injected (4 ml). After 10 minutes of QDs growth, the lecithin solution was added to the crude $CsPbBr_3$ solution (4 ml). After 1 minute, the solution was concentrated on a rotavap at room temperature. The QDs were precipitated from the concentrated solution (5 ml) by an excess of acetone (12 ml). The obtained pellet was dissolved in 5 ml of toluene.

**9.4 nm LHP QDs.** The filtered hexane (24 ml) is mixed with PbBr2 stock solution (8 ml). Under heavy stirring, a 0.02 M Cs-DOPA stock solution was injected (4 ml). After 40 minutes of QDs growth, the lecithin solution was added to the crude CsPbBr3 solution (4 ml). After 1 minute, the QDs were precipitated by an excess of acetone (114 ml). The obtained pellet was dissolved in 5 ml of toluene.

**10.8 nm LHP QDs.** The filtered hexane (10 ml) is mixed with $PbBr_2$ stock solution (8 ml). Under heavy stirring, a 0.02M Cs-DOPA stock solution was injected (4). After 60 minutes of QDs growth the lecithin solution was added to the crude $CsPbBr_3$ solution (4 ml). After 1 minute, the obtained QDs were precipitated by an excess of acetone (66 ml). The obtained pellet was dissolved in 5 ml of toluene.

**18.4 nm LHP QDs.** Under heavy stirring, a 0.2 M Cs-DOPA stock solution (0.2 ml) was injected into 4ml of PbBr2 stock solution. After 30 minutes of QDs growth 10 ul of oleic acid and oleylamine were added. Then 2 ml of 0.02M Cs-DOPA stock solution were slowly injected with 2 ml/h rate. After injection was completed, the lecithin solution was added to the crude CsPbBr3 solution (2 ml). After 1 minute, the obtained QDs were precipitated by an excess of acetone (22 ml). The obtained pellet was dissolved in 5 ml of toluene.

**15 nm LHP QDs.** Detailed synthesis procedure of 15 nm LHP QDs, along with their structural and optical characterization can be found in our previous paper.[3]

**3.9 nm CdSe.** Sample preparation detail along with their characterization can be found in our previous paper.[4]

## 1.2. Transmission Electron Microscope (TEM) characterization

TEM images were collected using a JEOL JEM-1400 Plus operated at 120 kV or Hitachi HD-2700. The samples were prepared by placing a drop of diluted QDs solution on coated Cu TEM grids.

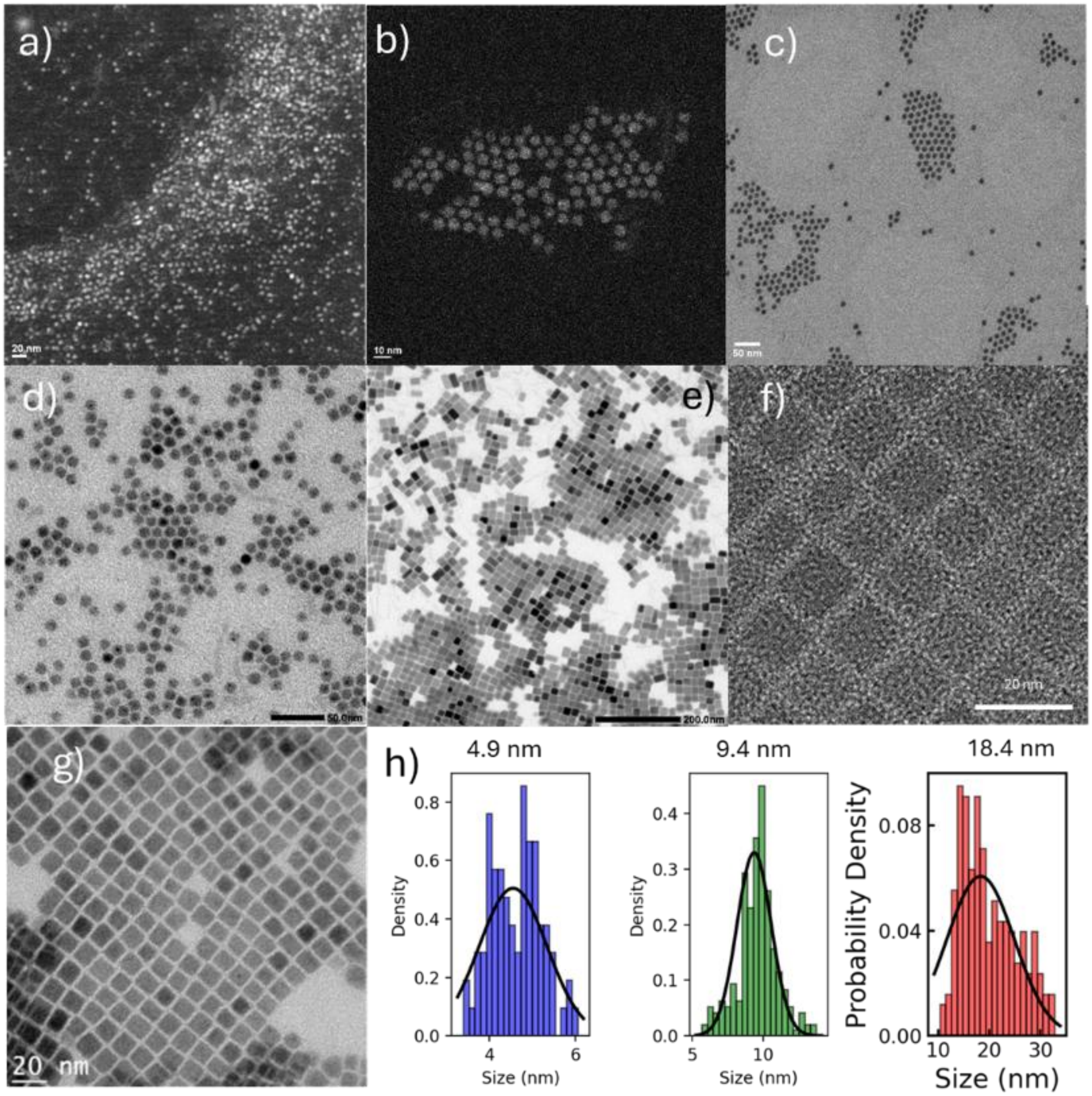

**Figure S1:** STEM image of the (a) 4.9 and (b) 5.9 nm $CsPbBr_3$ QDs prepared by Kovalenko. TEM images of (c) 9.4 nm, (d) 10.8 and (e) 18.4 nm large $CsPbBr_3$ QDs prepared by Kovalenko. TEM image of (f) 15 nm $CsPbBr_3$ QDs, (g) $CsPbI_3$ QDs prepared by Kambhampati. (h) Histogram of diameters for 4.9 nm,9.4 nm and 18.4 nm P QD.

### 1.3. Linear Spectroscopy characterization

Optical characterizations were performed at ambient conditions. UV-Vis absorption spectra of colloidal QDs were collected using a Jasco V670 spectrometer in transmission mode. The NC concentrations were determined from the absorption spectra using the absorption coefficient reported by Maes et al.[5] For the measurements QDs solutions were diluted down to 20-50 µg/mL. Zwitterion-capped QDs were dispersed in either hexane or toluene.

A Fluorolog iHR 320 Horiba Jobin Yvon spectrofluorometer equipped with a PMT detector was used to acquire steady-state PL spectra. NC solutions were measured in the same dilutions and solvents as the absorption measurements.

Photoluminescence quantum yield (PLQY): Absolute PL QYs of solutions were measured with a Hamamatsu C13534 Quantaurus-QY Plus UV-NIR absolute PL quantum yield spectrometer. The same solutions that were used to measure PL were also used to measure QY.

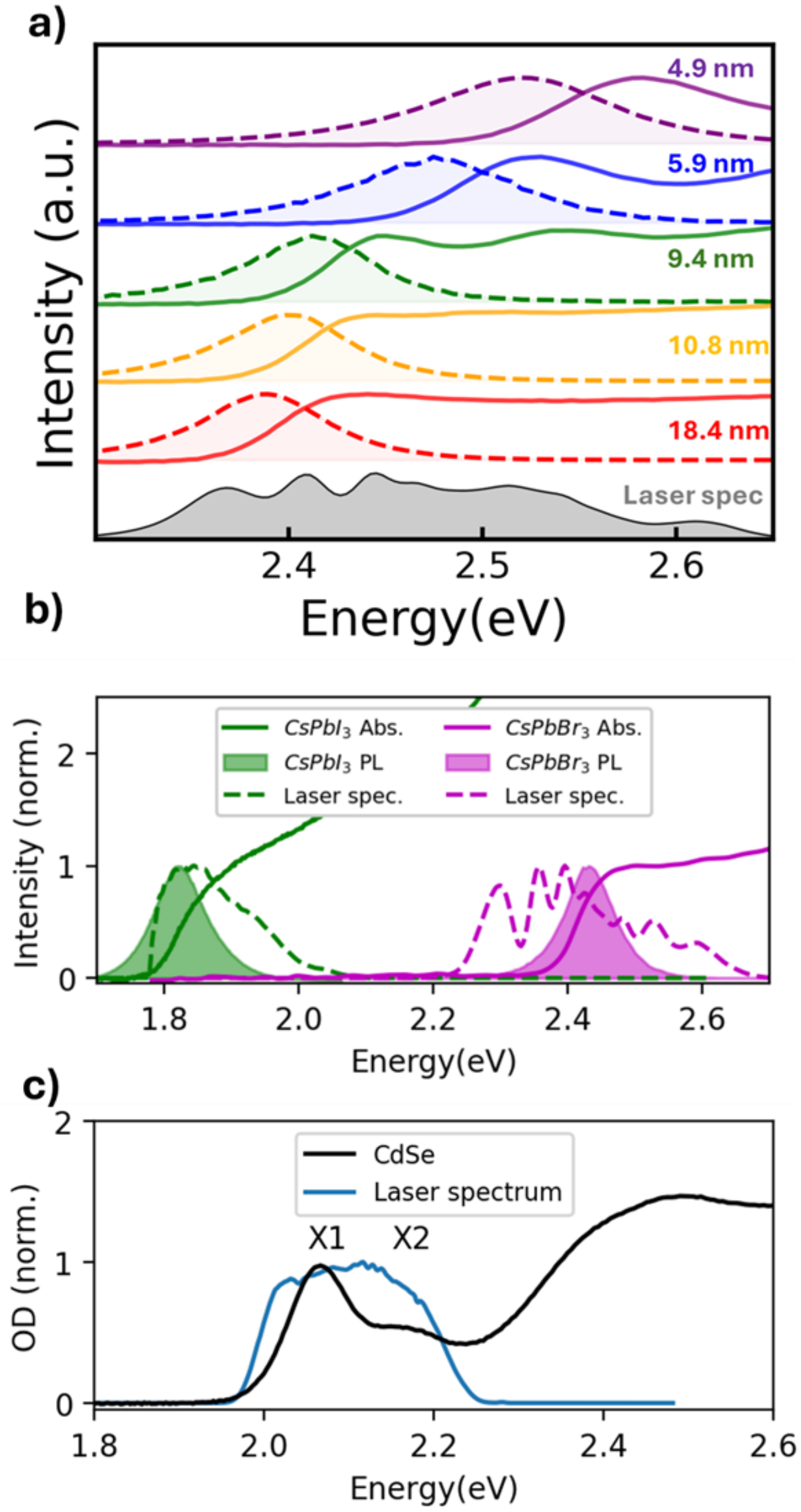

**Figure S2:** a) Linear absorption spectra (solid lines), photoluminescence spectra (dashed line with shaded area) of 18.4 nm (red), 10.8 nm (orange), 9.4 nm (green), 5.9 nm (blue) and 4.9 nm (violet) QDs. Also shown in bottom panel is laser spectra (grey). Linear spectra and laser spectra used for this study for (b) 15 nm LHP QDs and (c) 3.9 nm CdSe QD.

## 2. Coherent Multi-Dimensional Spectroscopy

The Coherent Multi-Dimensional Spectroscopy (CMDS) instrument was previously described in detail [3,4,6-17]. The femtosecond light source was an Ar filled hollow core fiber (HCF) which produced spectrally broadened pulses based upon the 500 nm output from a 120 fs optical parametric amplifier. The coherent pump pulse train was created by acousto-optic modulators. The CMDS experiment was conducted in the pump/probe geometry. **Figure S3** illustrates the CMDS experiment and signals.

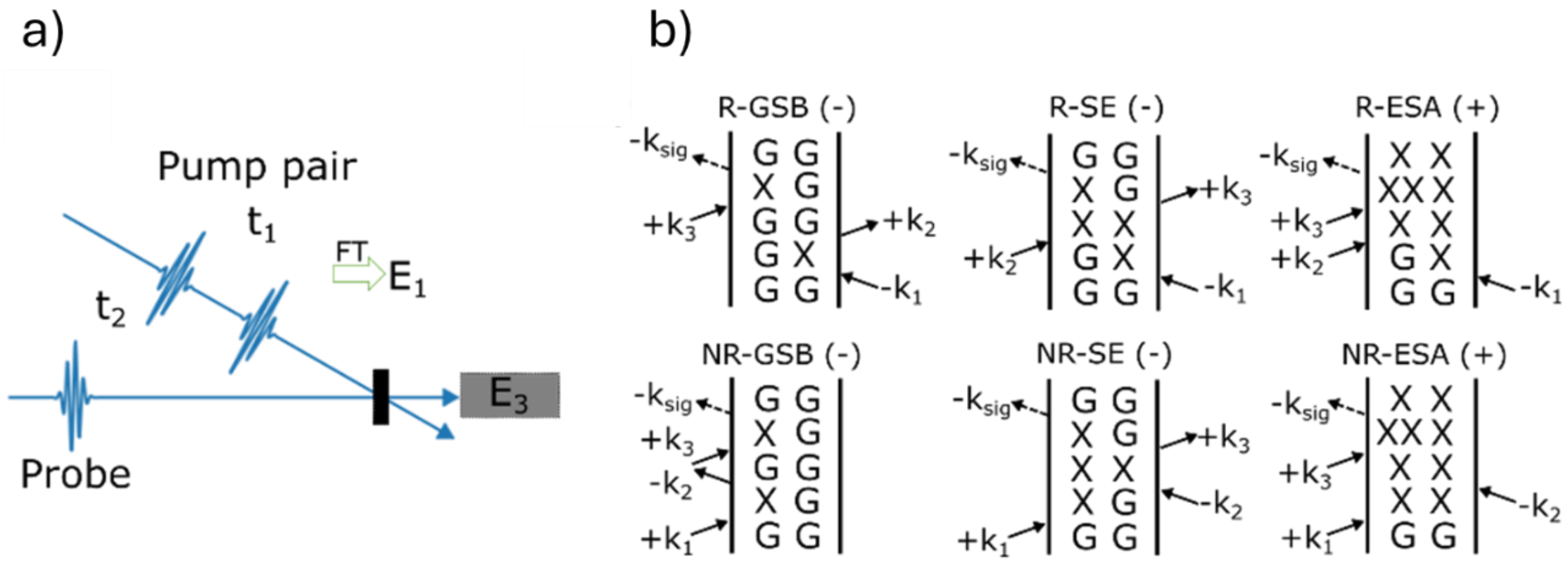

**Figure S3:** Illustration of CMDS experiment, Feynman diagrams for the six response functions.

## 3. Pulse characterization

Pulses are suitably manipulated to obtain near-transform limited pulses at the sample position. The pulses are characterized using a home built all-reflective dispersion-free transient-grating frequency resolved optical gating (TG-FROG). Pulse durations are determined by fitting a Gaussian pulse shape to the time marginal of the FROG trace. Panel a, b and c show pulses used to conduct 2DES at different wavelength ranges for the 2DES experiments of $CsPbBr_3$, CdSe and $CsPbI_3$ QDs, respectively. The temporal duration of the pulse is estimated to be close to 10 fs (shown in panel a), 15 fs (panel b) and 13-15 fs (panel c), respectively.

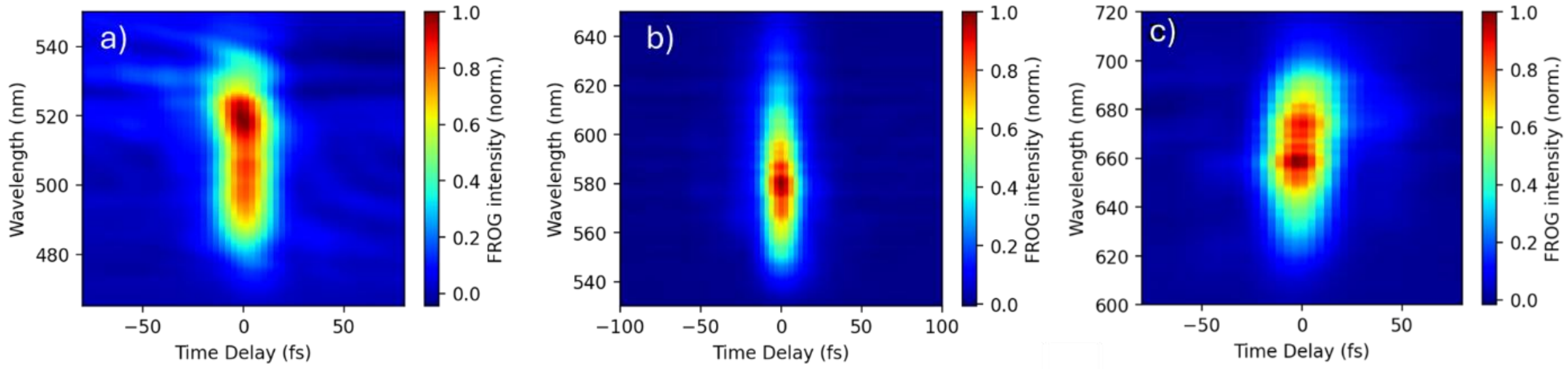

**Figure S4:** Typical TG-FROG trace used for 2DES experiments on a) $CsPbBr_3$ PQDs, b) CdSe QDs, and c) $CsPbI_3$ PQDs.

## 4. Sample handling and solvent response.

All the experiments were carried out at room temperature with samples dispersed in toluene. For the 2D experiments, the optical density of the corresponding sample was 0.2 at the band-edge, and the sample was constantly flowed in a 200 µm thick glass cuvette using a peristaltic pump. Pulse energies were≃10 nJ/pulse for all three pulses. Several repetitions of the experiments were carried out on different days. Pure solvent runs were executed to probe the non-resonant response arising from the solvent and glass cuvette windows. The kinetic transients for the X1-X1 and X2-X1 peaks of both solvent and P NC is shown in **fig. S5, a.** The transients plotted in dashed lines are for toluene and solid line shows response of 18.4 nm P NC. The same transients without re- normalization is shown in **Fig. S5,b** , showing that the solvent response is not significant compared the resonant sample response.

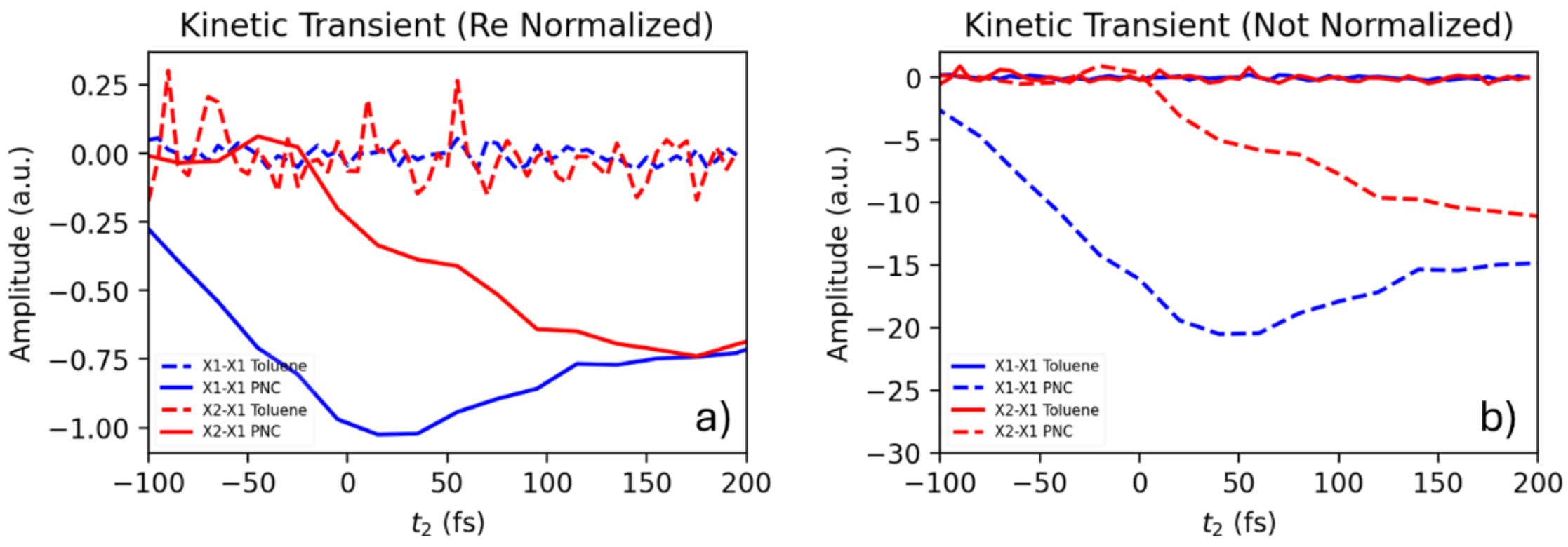

**Figure S5:** a) Kinetic transients from 18.4 nm P NC and toluene for the X1-X1 and X2-X1 peaks, re-normalized to show solvent response. b) Same transients as in panel (a) without renormalization.

**5. Handling early time perturbed free induction decay signals**

One of the main problems in nonlinear spectroscopy is the presence of early time signals, even at negative population times. These signals arise from Perturbed Free Induction Decay (PFID). We have discussed in detail the PFID signals in CdSe and CsPbBr3 NC QD. A correction of PFID signals is essential to careful signal analysis at early time[17].

<u>Negative population time artifacts</u>

When time delay $t_2$ is negative, the order of the pulse interactions is reversed. When $t_2 < 0$ and $|t_2| < t_1$, pulse $k_3$ arrives between pulse $k_1$ and $k_2$. The DSFD in **fig. S3,b** are reformed into the DSFD in **fig. S6,a**. The rephasing pathways are still rephasing pathways, though the order of the pulse interactions is changed. The non-rephasing pathways become two-quantum pathways.

In the context of two pulse pump-probe transient absorption spectra, the reversed pulse ordering signal is often referred to as perturbed free-induced decay (PFID)[18-20] , as the pump pulse arrives after the probe and perturbs the free induction decay signal being measured by the probe. **Fig. S6,b** depicts PFID in the context of the three pulse 2DE experiment when $t_1$ = 80 fs and $t_2$ = -50 fs. When the pulse ordering is [$k_1$, $k_2$, $k_3$], pulse $k_1$ initiates the free induction decay and pulse $k_2$ then excites the system into a population state (blue). When the pulse ordering is [$k_1$, $k_3$, $k_2$], pulse $k_1$ initiates the free induction decay and pulse $k_3$ prematurely interrupts the oscillating electronic coherence so the signal amplitude for $t_1$ = 30 fs is recorded instead of $t_1$ = 80 fs. Pulse $k_2$ then probes the system and triggers the emission of the third order polarization $k_{sig}$.

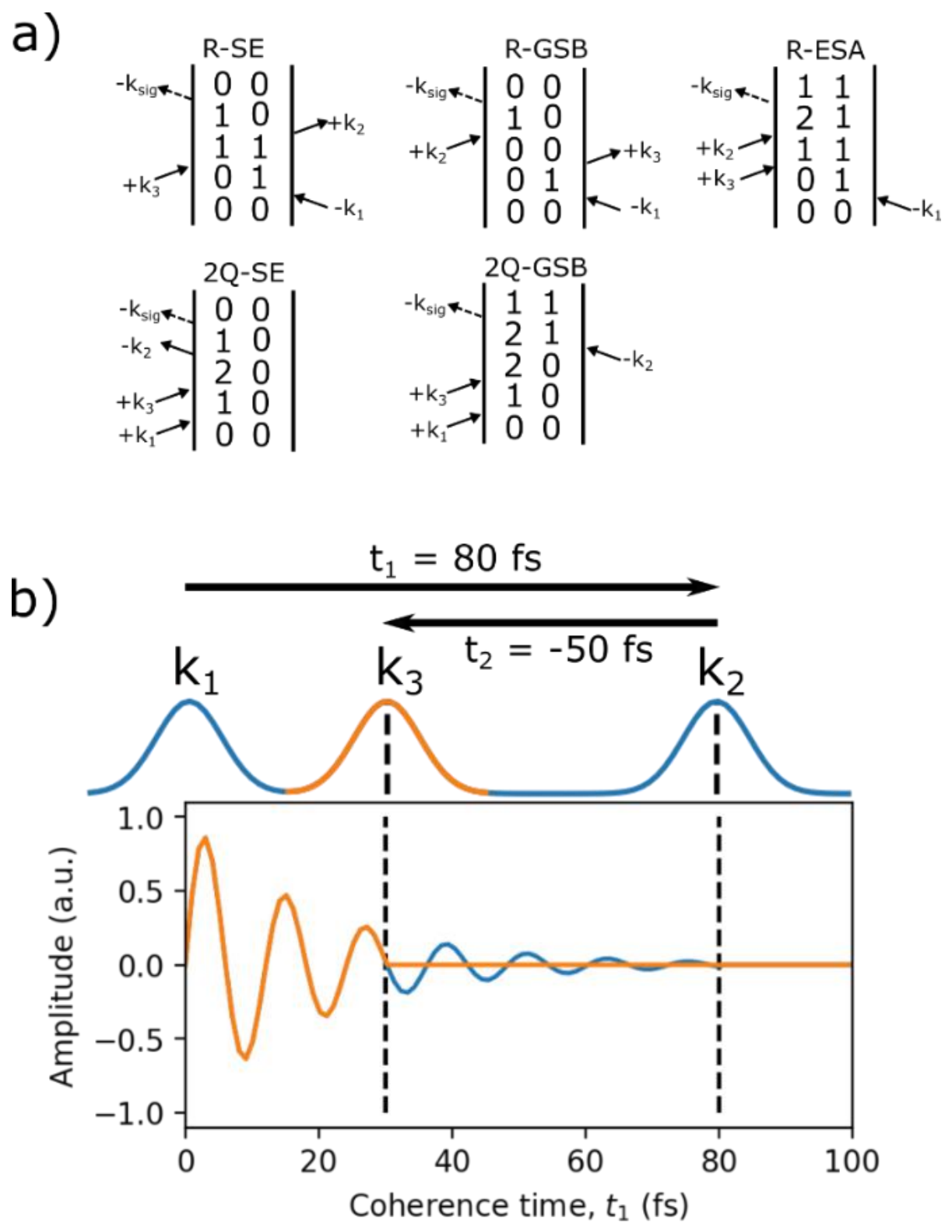

**Figure S6:** a) DSFD resulting from reversing the order of pulses 2 and 3. b) Perturbed free induction decay (PFID) from reversed pulse order, corresponding to $t_1$ = 80 fs and $t_2$ = -50 fs.

When $t_2 < 0$ and $|t_2| > t_1$, pulse $k_3$ arrives before pulse $k_1$ and $k_2$. The DSFD in **fig. S3,b** are reformed into the DSFD in **fig. S7,a**. The rephasing signal $-k_1 + k_2 + k_3$ becomes the non-rephasing signal $+ k_3 - k_1 + k_2$ and the non-rephasing signal $k_1 - k_2 + k_3$ becomes the two-quantum signal $k_3 + k_1 - k_2$. In the reversed pulse order, there would be a time delay between the emitted third-order polarization $k_{sig}$ and the local oscillator $k_3$. This time delay creates spectral fringes at negative $t_2 << 0$. As the fringes at negative time delays are significantly narrower than the desired 2DE lineshapes, the fringes can be smoothed by Fourier filtering without significantly distorting the 2DE lineshapes[21,22] .

After Fourier filtering, negative time signals still appear at -50 fs < $t_2$ < 0. These negative time signals artificially shift the X1-X1 peak to earlier population times, as is apparent in the kinetic transients is **fig. S7 b,c**.

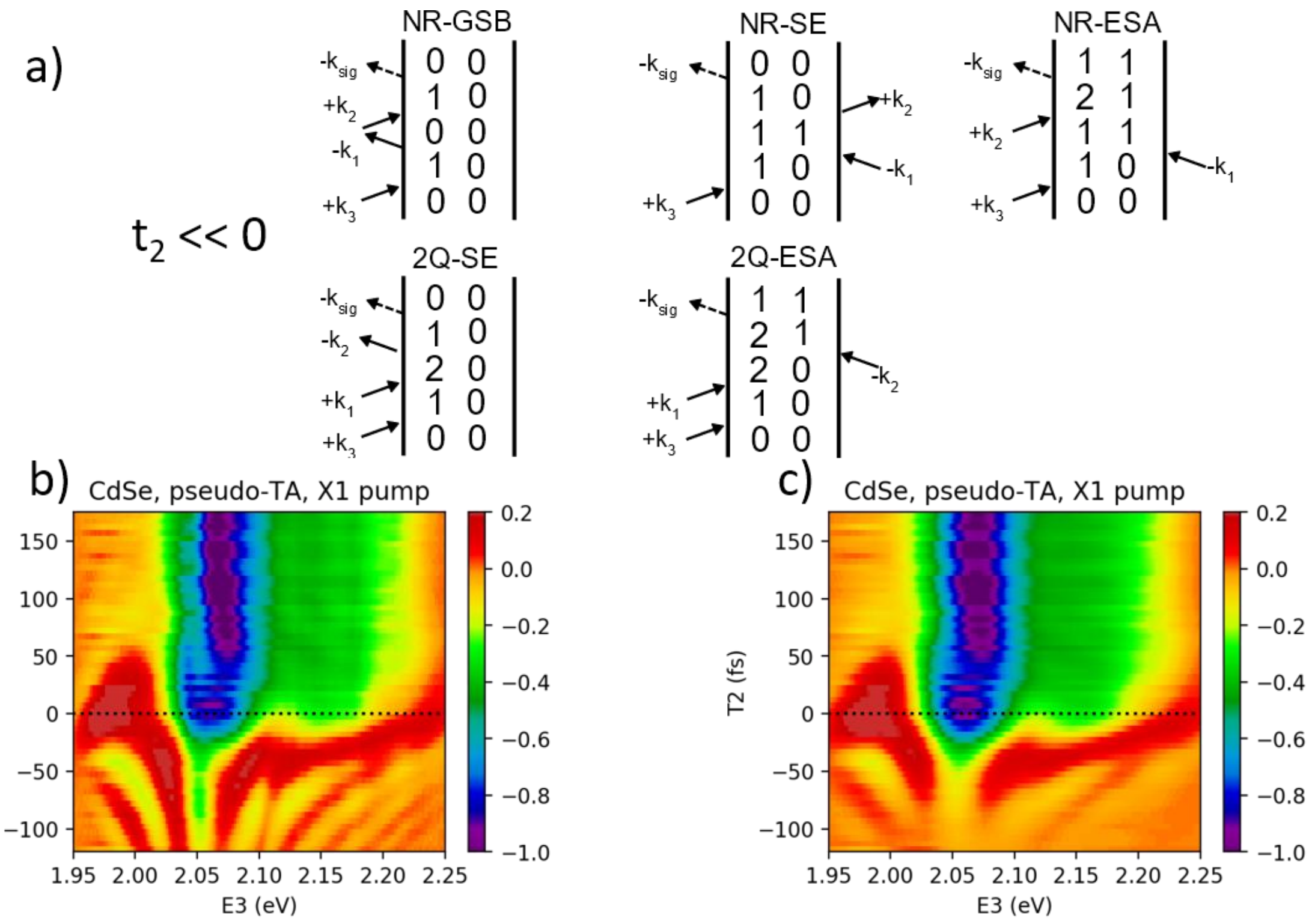

**Figure S7:** a) Example DSFD when $t_2 > 0$ and $t_2 < 0$. b) 2DE spectrum before Fourier filtering. c) 2DE spectrum after Fourier filtering. Pseudo-TA spectra generating with a 5 meV radius.

## 6. CMDS spectra, pseudo–Transient Absorption (TA) projection and peak shifts

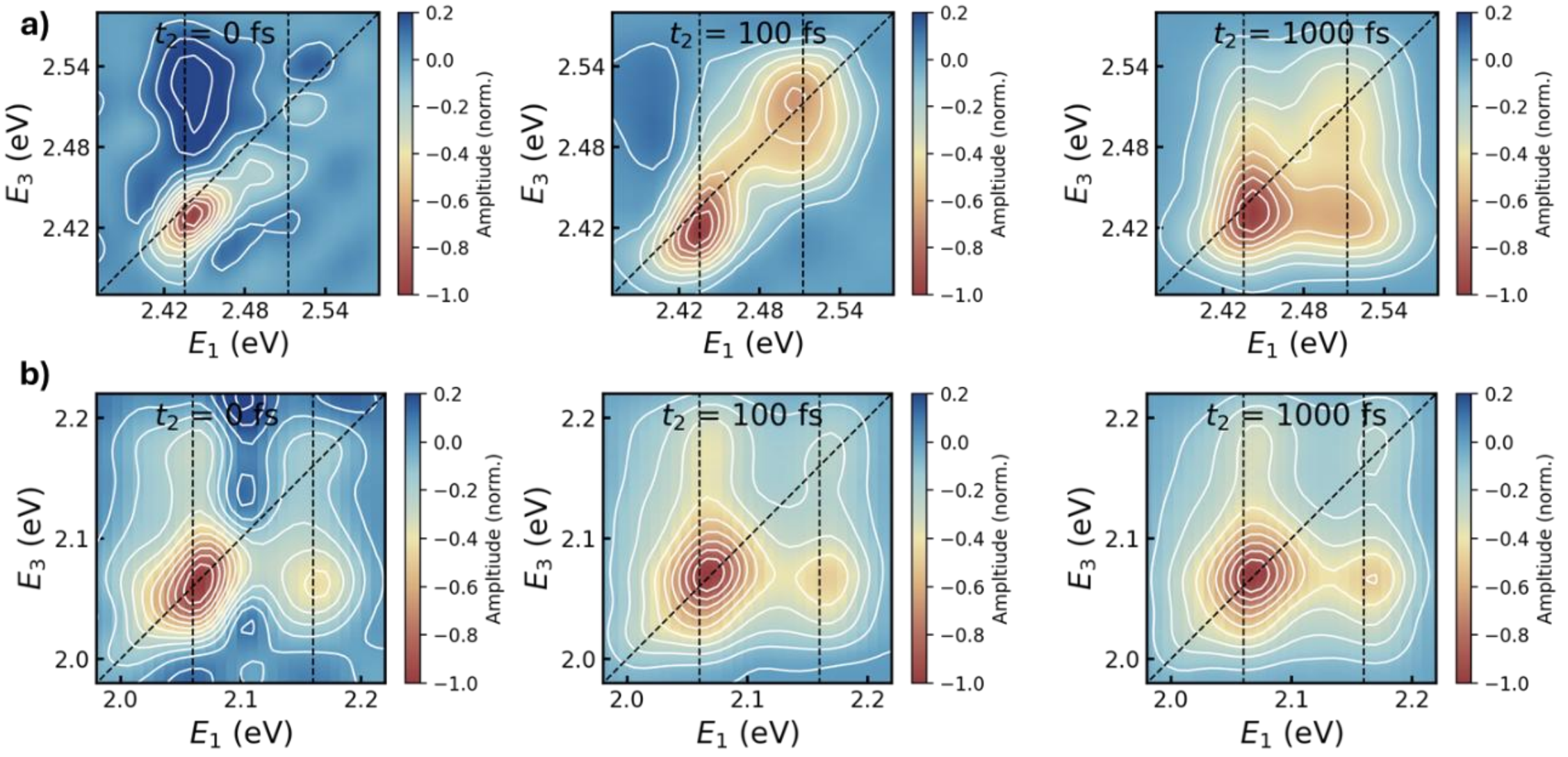

**Figure S8:** Same CMDS spectra as shown in Figure 3 of the main text but plotted with an expanded positive amplitude scale to reveal additional background signals at early times. These features are attributed to excited-state absorption processes, along with the prominent excitonic signatures.

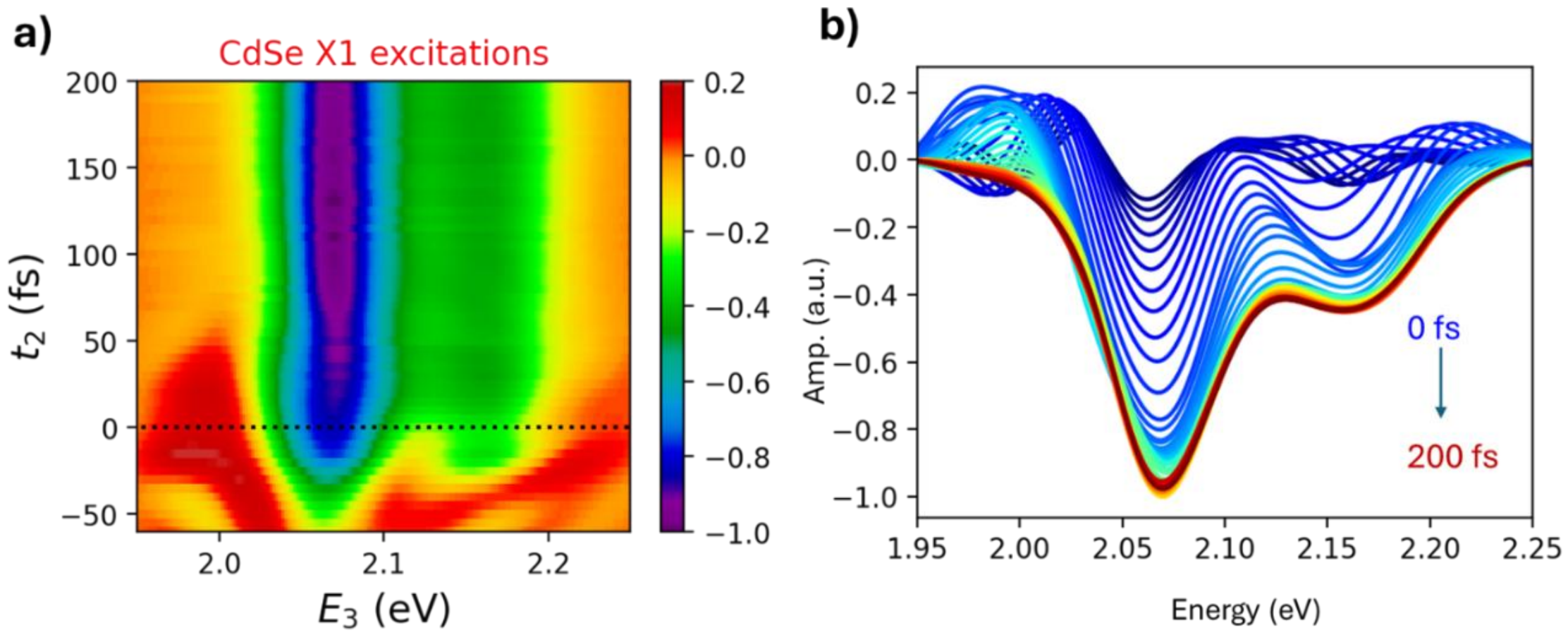

**Figure S9:** Projection of the CMDS spectra along $E_3$ energy axis (pseudo-TA slice) and spectral projection at different population time at X1 excitations showing fine structure relaxation for 3.9 nm CdSe QDs. (a) pseudo TA spectral projection obtained from CMDS spectra after X1 excitation, b) spectral slices at different population time projected on energy axis for same CdSe QDs.

**Extraction of $X_1$ peak shift dynamics:** To quantify the time-dependent spectral shift of $X_1$ excitonic features, we analyzed the evolution of slices from CMDS dataset S ($E_1$, $t_2$, $E_3$) as a function of the population time delay ($t_2$). A 40 meV window centered at the $X_1$ band edge excitonic transition was selected along $E_1$ excitation axis. The data within this window were integrated to yield a 2D matrix S ($t_2$, $E_3$), representing the time-resolved spectral evolution along the emission axis $E_3$. Each spectrum was baseline-corrected and normalized to its global maximum. For each $t_2$, the $X_1$ peak was identified as the local minimum (bleach maximum) via Gaussian fitting. The extracted peak energies were then referenced to the average of the first few time points to obtain transient peak shift. Temporal peak shift was modeled further based on a convolution-based fitting approach to capture the dynamics.

## 7. Diagonal and Anti Diagonal spectral projection

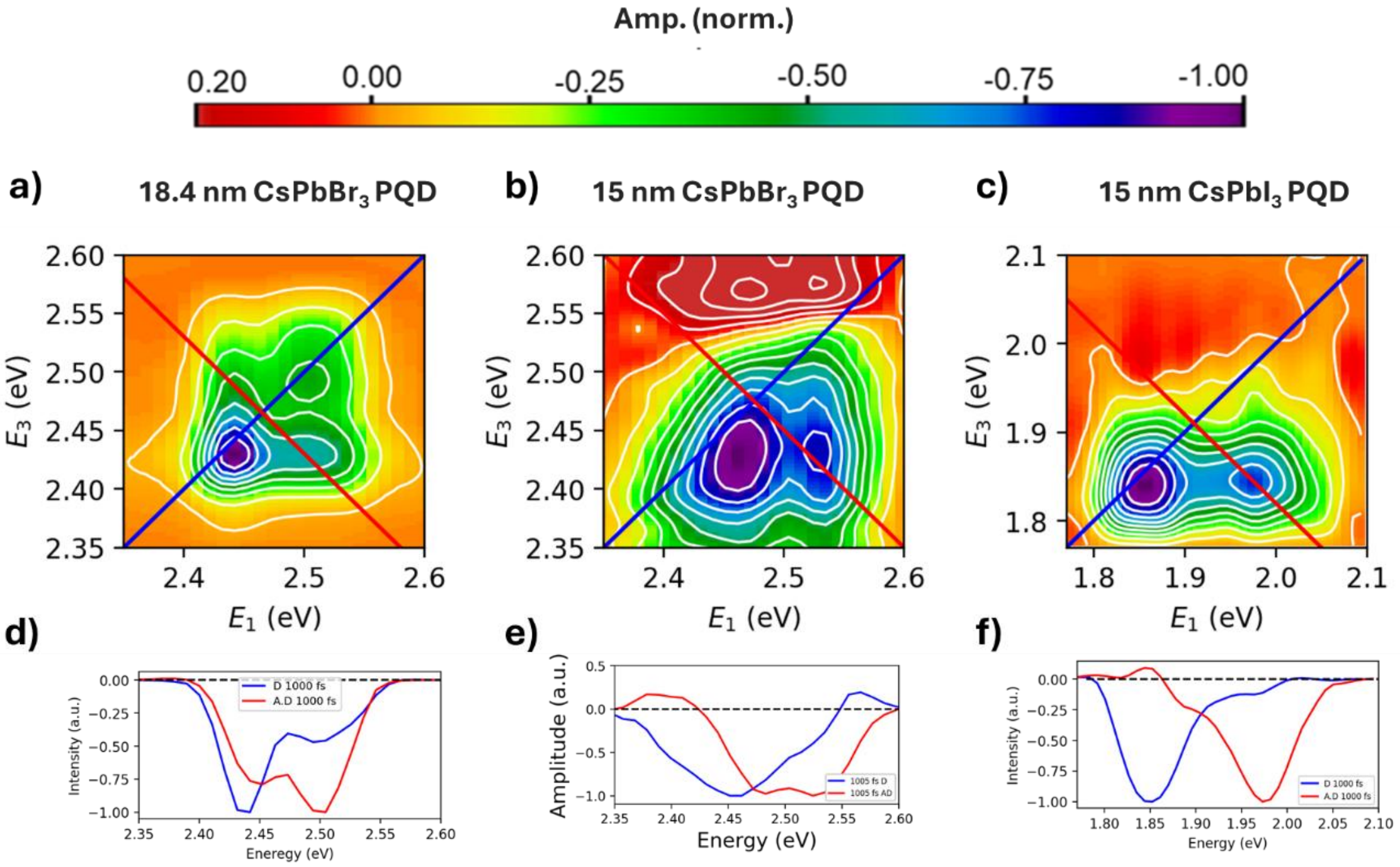

**Figure S10:** Representative data showing energy splittings along the diagonal and anti-diagonal spectral projections. The diagonal spectra reflect the excitation spectrum of the exciton manifold, while the anti-diagonal spectra reveal splittings due to couplings between states. Data is shown for 18.4 nm, 15 nm $CsPbBr_3$, and 15 nm $CsPbI_3$ QDs, where the 18.4 nm LHP QDs were synthesized

by Kovalenko, and the remaining sizes were synthesized by Kambhampati. The top panel presents CMDS spectra at late times, with corresponding diagonal and anti-diagonal projections shown in the bottom panel for different LHP QDs.

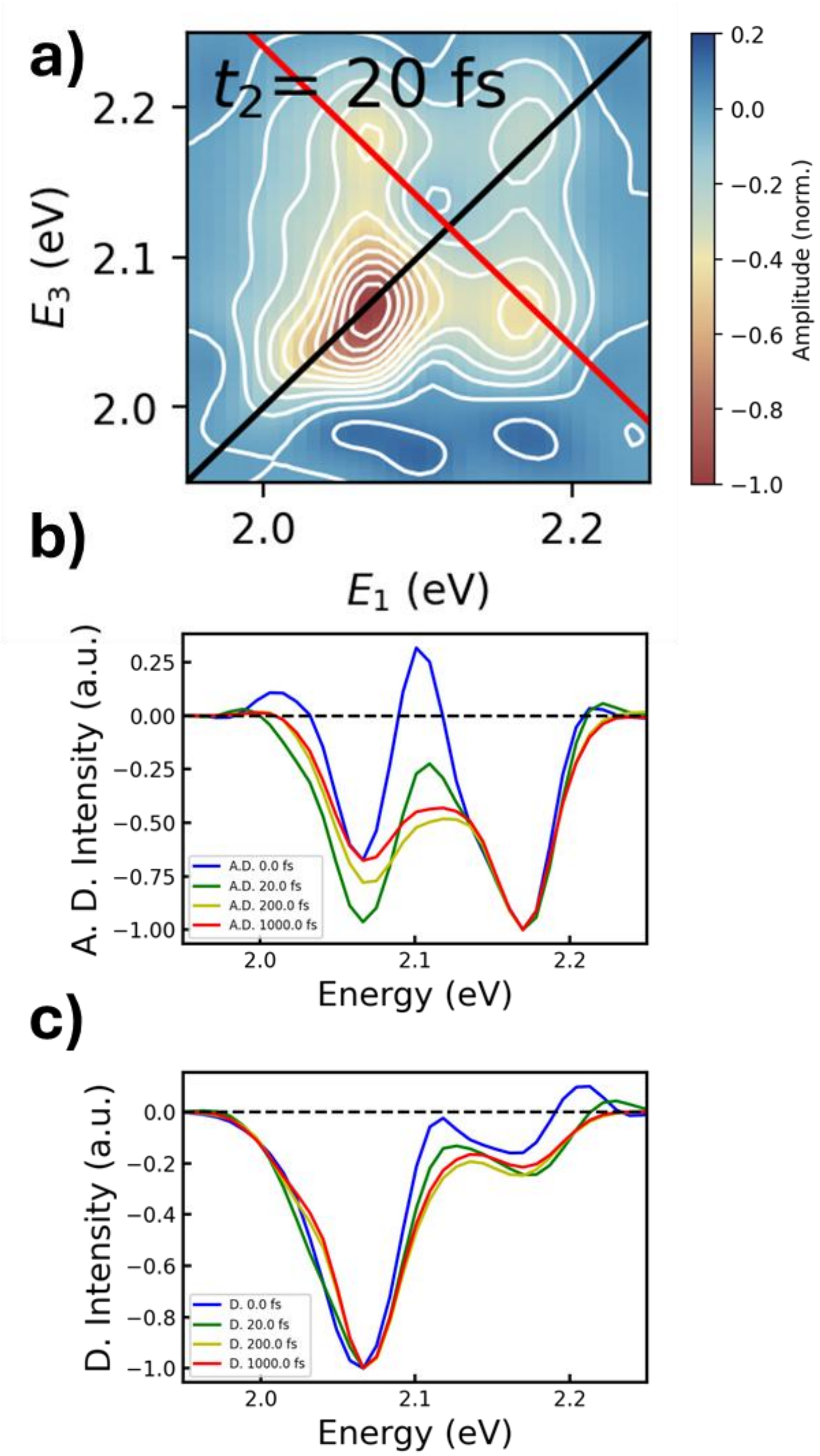

**Figure S11:** Representative data showing four static peaks always present, both along the diagonal and off-diagonal. (a) CMDS data of 3.9 nm CdSe QDs at early population time, (b) anti-diagonal projection, and (c) diagonal projection for different population times.

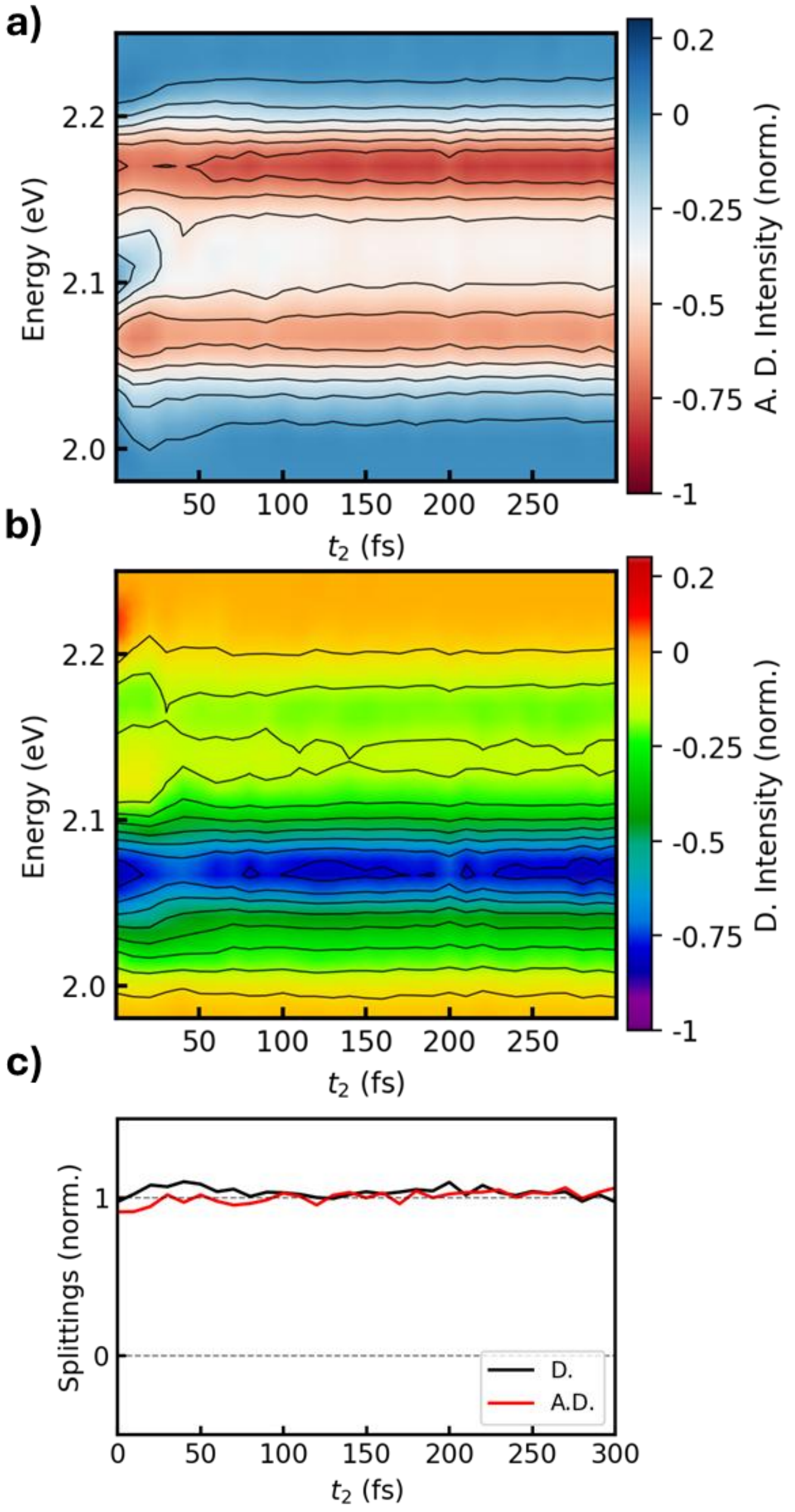

**Figure S12:** CMDS shows that couplings in CdSe are static unlike dynamic coupling of PQDs. a) anti diagonal and b) diagonal spectral dynamics showing couplings of X1 and X2 peaks. c) The AD and D normalized peak splittings vs time.

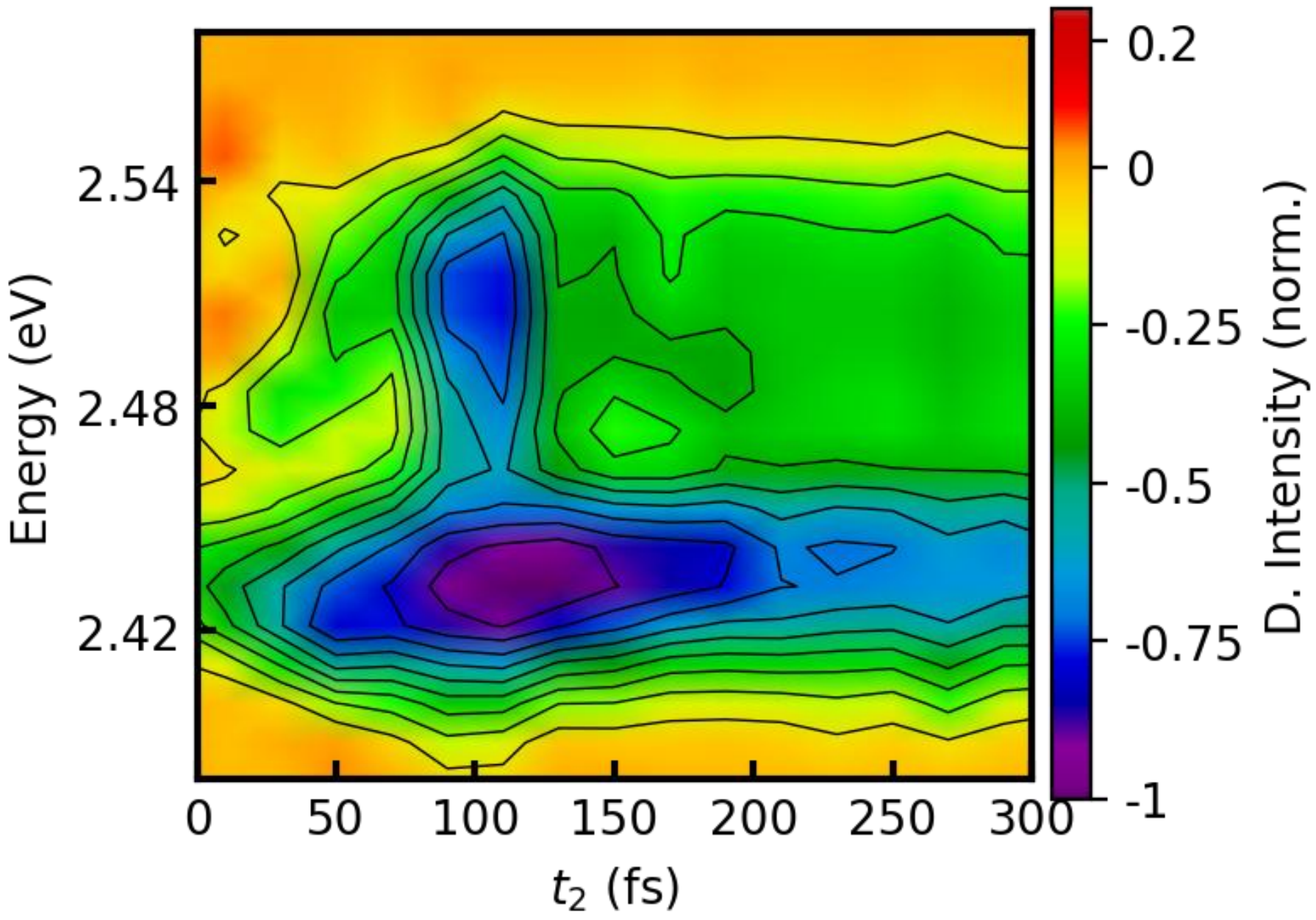

**Figure S13:** Diagonal spectral dynamics for 18.4 nm PQD showing time-dependence of diagonal coupling.

**8. Size dependence of other observables**

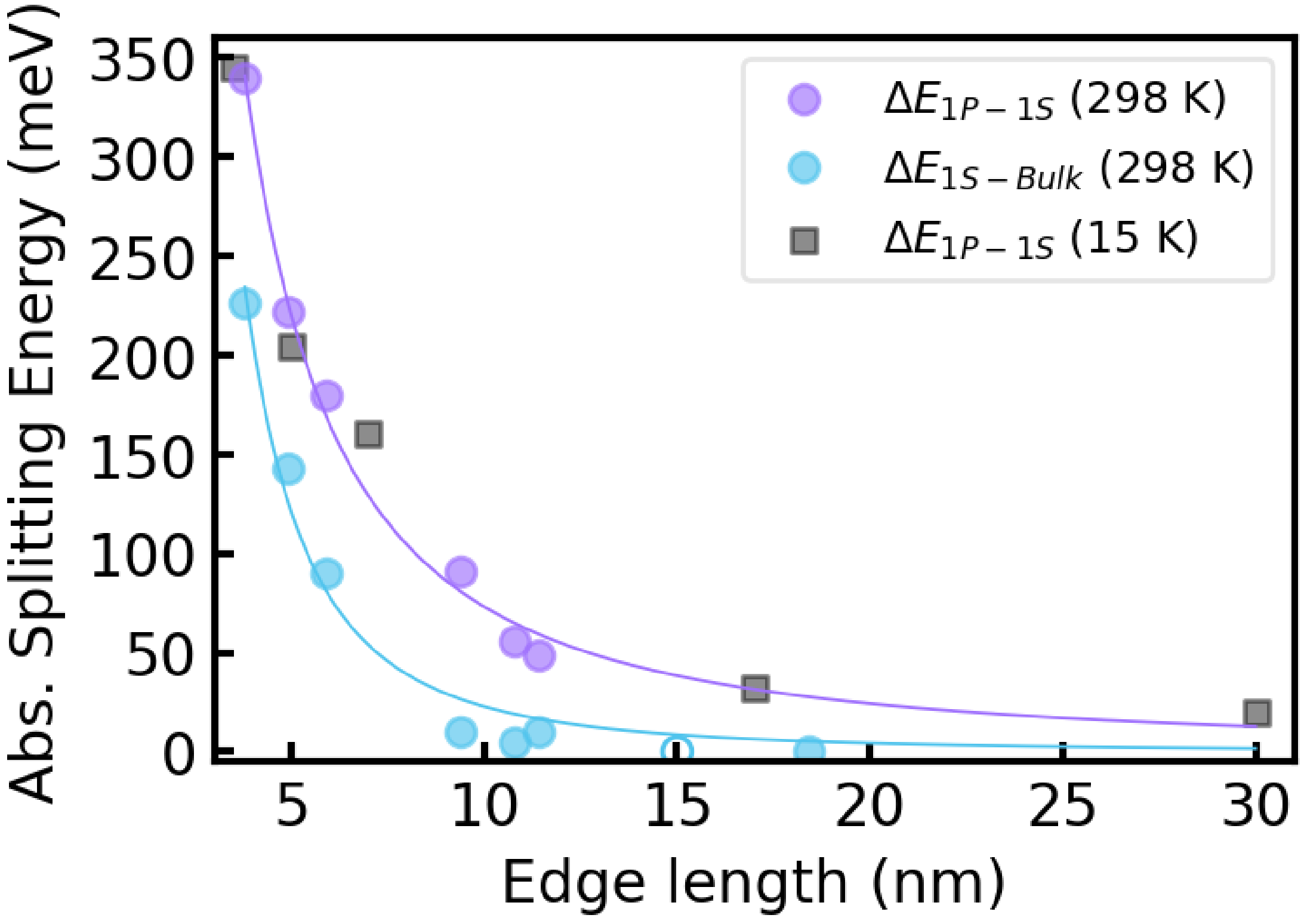

**Figure S14:** Extended energy scale version of Figure 5C from the main text. This plot highlights the energy of the band-edge exciton relative to the bulk reference and illustrates the 1P–1S energy separation arising from quantum confinement at room temperature and 15K.

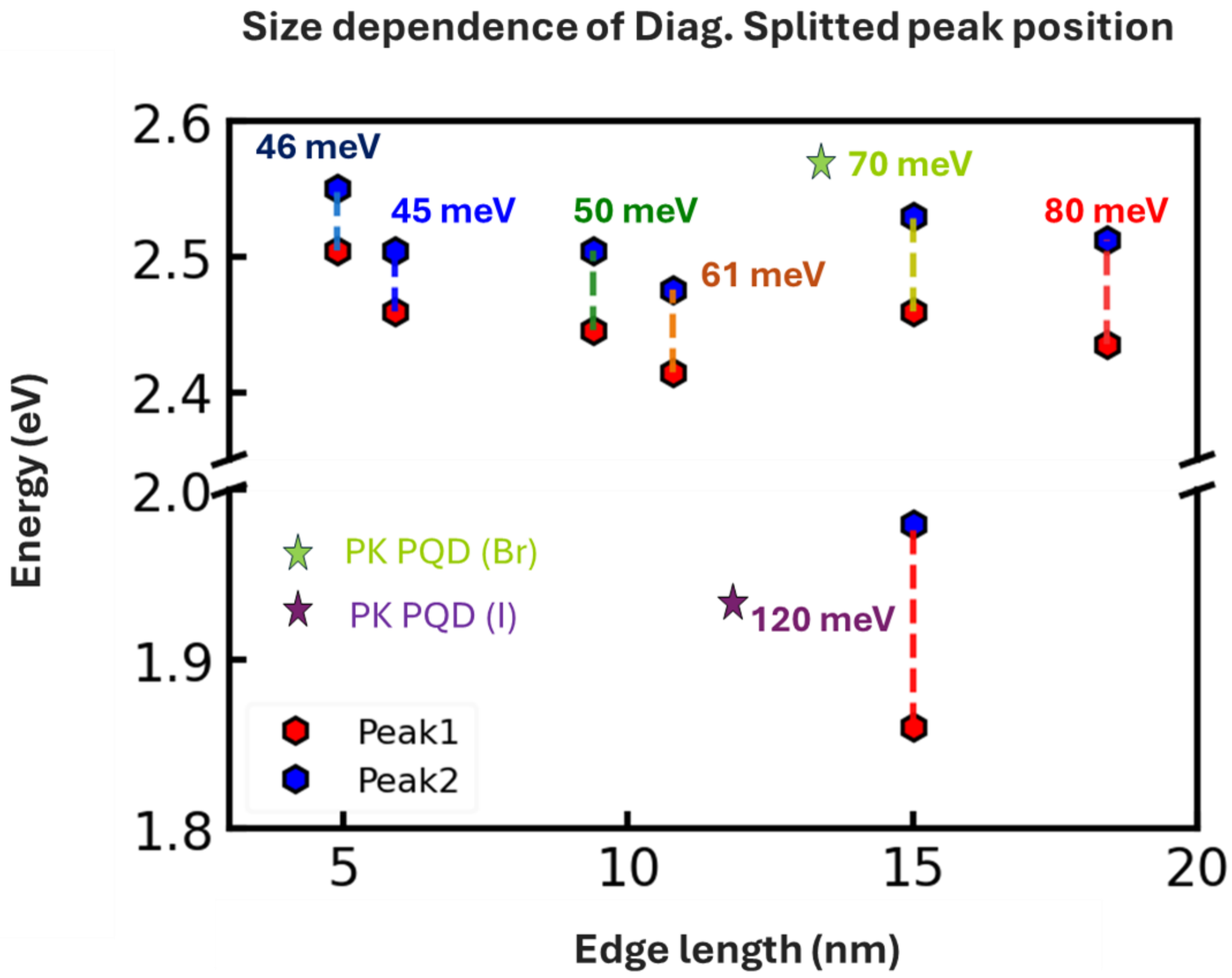

**Figure S15:** Split peak position along diagonal energy projection at late times from CMDS spectra, following the completion of lattice structural dynamics, reflecting exciton manifolds.

## 9. Polaron-induced dynamic quantum confinement in a perovskite quantum dot: Hamiltonian formalism, dynamics, and connection to CMDS observables

### 9.1 Physical overview and experimental context

We study a single ∼ 20 nm all-inorganic perovskite quantum dot (PQD) at room temperature, photoexcited into its bright band-edge exciton state. In such a polar, soft lattice, strong exciton–phonon coupling can form a polaron, a composite quasiparticle of the exciton and a coherent lattice deformation.

In our case, the phonon bath is overdamped and liquid-like, as revealed by the low-frequency Raman spectrum. This leads to a time-dependent polaron coordinate $Q(t)$ that:

1. **Shifts** both excitonic states in parallel (blue drift of ∼+5 meV).
2. **Forms a confinement well** that quantizes the exciton center-of-mass (COM) motion, producing a higher-energy exciton $X_2$ ∼50 meV above the band-edge $X_1$.
3. **Turns on interstate coupling** between $X_1$ and $X_2$ after the well forms, producing delayed growth of anti-diagonal (AD) peaks in CMDS spectra.

The combined effects produce three experimental hallmarks in absorptive CMDS:

1. **Diagonal peak:** one at $t_2$=0 fs; two peaks by ∼150 fs; both shifting in parallel.
2. **Anti-diagonal structure:** absent before ∼100 fs, mid-ridge at ∼150 fs, then two well-separated lobes by 300 fs.
3. **Amplitude growth:** diagonal signals rise by a factor ∼2.5 over the polaron build-up time, consistent with ordering of unit-cell dipoles.

### 9.2 The Hamilton formulation

We start from a compact Hamiltonian that captures the exciton-phonon physics and its connection to experimental observables:

#### 9.2.1 Electronic subspace

We take two bright excitonic basis states:

$|1\rangle$: the **band-edge** exciton (delocalized over the QD).

$|2\rangle$ : the **polaron-confined COM** state (becomes bright only after confinement).

The electronic Hamiltonian depends explicitly on the polaron coordinate $Q$:

$$H_{el}(Q) = \begin{pmatrix} E_1^0 + \alpha Q & V_{\mathrm{mix}}(Q) \\ V_{\mathrm{mix}}(Q) & E_1^0 + \Delta_{\mathrm{COM}}(Q) + \alpha Q \end{pmatrix}$$

Where $\alpha Q$: common linear shift from the polaron's dielectric effect → the parallel blueshift of both states.

$\Delta_{\mathrm{COM}}(Q)$: COM quantization energy in the polaron well:

$$\Delta_{\mathrm{COM}}(Q) \equiv \frac{\hbar^2 \pi^2}{2 M_X R_{\mathrm{pol}}^2(Q)} + \delta V(Q)$$

Here $R_{pol}(Q)$: is the confinement radius and $\delta V(Q)$ accounts for finite-well-depth corrections.

$V_{mix}(Q)$: interstate coupling, rising as the well forms-responsible for delayed AD growth.

**9.2.2 Phonon bath and system-bath coupling**

We describe the lattice using a reaction-coordinate approach, equivalent to a Holstein/Fröhlich model mapped to an effective coordinate $Q$ coupled to a bath of harmonic oscillators:

$$H_{\mathrm{ph}} = \frac{P_Q^2}{2M_Q} + \frac{1}{2} K_Q Q^2 + \sum_k \left( \frac{p_k^2}{2m_k} + \frac{1}{2} m_k \omega_k^2 q_k^2 \right)$$

$$H_{\mathrm{coup}} = -Q[g_0 \,\widehat{n_X} + g_{12}(|1\rangle\langle 2| + |2\rangle\langle 1|)] - Q \sum_k c_k q_k$$

Where: $\widehat{n_X} = (|1\rangle\langle 1| + |2\rangle\langle 2|$ is the total exciton number, $g_0$ is coupling that shifts both exciton energies in parallel, $g_{12}$ is coupling that modulates $V_{mix}$ through the lattice, $c_k$ is couplings to residual bath mode.

### 9.2.3 Spectral density from Raman

The bath spectral density $J(\omega)$ is extracted from the low-frequency Raman susceptibility $\chi''_{\mathrm{Raman}}(\omega)$ :

$$J(\omega) = \frac{\pi}{2} \sum_k \frac{c_k^2}{m_k \omega_k} \, \delta(\omega - \omega_k) \; \propto \; \omega \, \chi''_{\mathrm{Raman}}(\omega)$$

The Raman spectrum is well described by a Drude-Lorentz (overdamped) form:

$$J(\omega) = \frac{2\lambda\gamma\omega}{\omega^2 + \gamma^2}$$

Where $\lambda$ is the reorganization energy and $\gamma = 1/\tau_{pol}$ is the inverse polaron build-up time.

From experiment: $\tau_{pol} \approx 150$ fs.

### 9.3 Polaron coordinate dynamics

Integrating out the bath gives a generalized Langevin equation for $Q(t)$:

$$\tau_{\mathrm{pol}} \, \dot{Q}(t) = -\big[Q(t) - Q_{\mathrm{eq}} \, \widehat{n_x}\big] + \zeta(t)$$

$$\langle \zeta(t) \, \zeta(0) \rangle \propto k_B T \, e^{-|t|/\tau_{\mathrm{pol}}}$$

After exciton creation $Q(t)$ rises exponentially:

$$Q(t_2) = Q_\infty \big(1 - e^{-t_2/\tau_{\mathrm{pol}}}\big)$$

### 9.4 How $Q(t)$ drives the CMDS observables

#### *9.4.1 Parallel blueshift:*

$$E_{1,2}(t_2) = E_{1,2}^0 + \alpha Q(t_2) \quad \Rightarrow \quad \Delta E_{X_1}(t_2) \to +5 \text{ meV}$$

Both $X_1$ and $X_2$ shift together.

9.4.2 Emergence of X2:

$$\Delta_{\mathrm{COM}}(t_2) \to 50 \text{ meV}$$

As $R_{pol} \to 4$ nm (diameter 8 nm), matching the extracted polaron geometry.

9.4.3 Delayed AD coupling:

$$V_{\text{mix}}(t_2) = V_\infty\left(1 - e^{-(t_2 - t_{\text{on}})/\tau_{\text{pol}}}\right)H(t_2 - t_{\text{on}})$$

With $t_{\text{on}} \approx 100$ fs → no AD before this; then mid-ridge two lobes by 300 fs.

**9.5 Extracted microscopic parameters**

From fits to Raman and CMDS data:

| **Parameter** | **Value** | **Physical significance** |
|---|---|---|
| $\tau_{pol}$ | 150±20 fs | Polaron build-up time from Raman and kinetics |
| $D_{pol}$ | $8 \pm 1$ nm | Confinement diameter from COM spacing |
| $V_\infty$ | ~70 meV | Well depth from COM ladder |
| $V_{mix,\infty}$ | 10-20 meV | Late-time interstate coupling |
| Blue drift | +5±1 meV | Dielectric screening change |
| $t_{\text{on}}$ | 100±20 fs | Delay before coupling onset |

**9.6 Full Second-Quantized Hamiltonian for the Polaron–Exciton System in a PQD**

We model the PQD as a finite spherical potential well (the polaron drop) that hosts a small number of discrete exciton center-of-mass (COM) states. The exciton couples linearly to a bath of optical phonons that produces a self-consistent polaron displacement field, shifting and distorting the exciton potential.

The total Hamiltonian is

$$\hat{H} = \widehat{H_{\text{ex}}} + \widehat{H_{\text{ph}}} + \widehat{H_{\text{ex-ph}}} + \widehat{H_{\text{LM}}}.$$

1. **Exciton sector ($\widehat{H_{\text{ex}}}$)**

We restrict to the lowest two COM eigenstate $|1\rangle$ and $|2\rangle$ in the single exciton manifold, with fermionic electron ($e^\dagger$, $e$) and hole ($h^\dagger$, $h$) operators implicitly bound into a composite bosonic exciton operator $X_j^\dagger$.

$$\widehat{H_{\text{ex}}} = \sum_{j=1}^{2} E_j^0 X_j^\dagger X_j + J_0\left(X_1^\dagger X_2 + X_2^\dagger X_1\right)$$

$E_j^0$: bare energies of COM states in the absence of a polaron well.

$J_0$ : bare interstate coupling (off-diagonal mixing from static disorder, exchange, etc.).

$X_j^\dagger$ : creates an exciton in COM state $j$ (bosonic at low density).

2. **Phonon Bath $\widehat{H_{\text{ph}}}$**

We represent the nuclear polarization field by a bath of harmonic modes:

$$\widehat{H_{\text{ph}}} = \sum_q \hbar\omega_q \; b_q^\dagger b_q$$

$b_q^\dagger$ : creates a phonon of branch/frequency $\omega_q$ and wavevector $q$.

In lead halide perovskites, the dominant modes are polar LO phonons with frequency $\omega_{\text{LO}}$ and strong Fröhlich coupling.

3. **Exciton-phonon coupling $\widehat{H_{\text{ex-ph}}}$**

The exciton couples to the collective nuclear displacement coordinate $Q \propto (b^\dagger + b)$.

For each exciton state:

$$\widehat{H_{\text{ex-ph}}} = \sum_{j=1}^{2} \sum_q g_{jq} \; X_j^\dagger X_j\left(b_q^\dagger + b_{-q}\right)$$

$g_{jq}$ : exciton-phonon coupling constant, dominated by the Fröhlich term

$$g_{jq} \propto \frac{1}{q}\left(\frac{1}{\varepsilon_\infty} - \frac{1}{\varepsilon_0}\right) F_j(q)$$

With $F_j(q)$ the form factor of COM state $j$ in the dot. This term is diagonal in $j$ : phonons shift energies but do not change COM index directly.

When traced over bath, this coupling yields a self-consistent polaron shift:

$$E_j(t_2) = E_j^0 + \alpha_j\, Q(t_2)$$

With $Q(t_2)$ the mean nuclear displacement along the polaron coordinate.

4. **Light-matter interaction $\widehat{H_{\mathrm{LM}}}$**

In the dipole approximation:

$$\widehat{H_{\mathrm{LM}}}(t) = -\sum_j \left[\mu_j E^{(+)}(t) X_j^\dagger + \mu_j^* E^{(-)}(t) X_j\right]$$

$\mu_j$: transition dipole moments for $|g\rangle \leftrightarrow |Xj\rangle$ .

$E^{(+)}(t)$, $E^{(-)}(t)$ : positive/negative frequency component of the laser field.

### 9.7 Matrix representation in the single-exciton manifold

If we write the Hamiltonian for the $|X_1\rangle$, $|X_2\rangle$ subspace, with a single effective phonon coordinate $Q$:

$$H_{el}(Q) = \begin{pmatrix} E_1^0 + \alpha_1 Q & J_0 + \beta \mathrm{Q} \\ J_0 + \beta \mathrm{Q} & E_2^0 + \alpha_2 Q \end{pmatrix}$$

$\alpha_j Q$: diagonal polaron shifts (common mode when $\alpha_1 \approx \alpha_2$).

$\beta Q$: possible phonon-modulated interstate coupling.

$E_2^0 - E_1^0 = \Delta_{\mathrm{COM}}^0$: bare COM energy spacing in absence of polaron.

The instantaneous D-axis splitting is

$$\Delta_D(t_2) = \sqrt{[\Delta^0_{\text{COM}} + (\alpha_2 - \alpha_1)Q(t_2)]^2 + 4[J_0 + \beta Q(t_2)]^2}$$

In our case, $J_0$ and $\beta Q$ are small at long $t_2$,so $\Delta_D \approx \Delta_{COM}$ measures the polaron-dressed COM spacing.

**9.7 Raman to spectral density to polaron dynamics**

**1. Raman data to phonon spectrum**

Experimental low-frequency Raman gives the imaginary part of the polarizability response $\chi''(\omega)$. For a set of broadened harmonic modes:

$$\chi''(\omega) \propto \sum_k \frac{\gamma_k\, \omega}{(\omega^2 - \omega_k^2)^2 + \gamma_k^2 \omega^2}$$

where $\gamma_k$ is the damping. In lead halide PQDs, the broad low-frequency tail is fit by a Drude–Lorentz term, indicating overdamped motion.

**2. Spectral density**

The exciton–phonon coupling enters nonlinear optical response via the bath spectral density $J(\omega)$, which for a Drude–Lorentz form is:

$$J(\omega) = \frac{2\lambda\gamma\, \omega}{\omega^2 + \gamma^2}$$

$\lambda$: reorganization energy (strength of coupling).

$\gamma^{-1} = \tau_{\text{pol}}$: polaron formation time.

**3. From $J(\omega)$ to $Q(t)$**

The mean displacement for a sudden electronic excitation is:

$$Q(t) = Q_\infty[1 - e^{-\gamma t}]$$

With $Q_\infty = \lambda/K$ and $K$ the mode force constant. At long $t_2$ ($t_2 \gg \tau_{\text{pol}}$), $Q(t_2)$ saturates and the splittings become time independent.

**9.8 COM Quantization and Well Depth Extraction**

The polaron forms a finite spherical well of radius $R_{\text{pol}}$ and depth $V_\infty$**.** The COM energies are solutions of the finite-well Schrödinger equations:

$$\left[-\frac{\hbar^2}{2M_X}\nabla^2 + V_{\text{pol}}(r)\right]\psi_{nl}(r) = E_{nl}\psi_{nl}(r)$$

With $V_{\text{pol}}(r) = -V_\infty \quad \text{for}\, r < R_{\text{pol}},\ and\ 0\ \text{outside}.$

The $n = 1,\ \mathrm{l} = 0\ corresponds\ to\ |X_1\rangle;\ n = 1,\ \mathrm{l} = 1\ \ corresponds\ to\ |X_2\rangle.$

For large $V_\infty$ , the COM spacing approaches the infinite-well form:

$$\Delta_{\text{COM}} \approx \frac{\hbar^2\left(\alpha_{n'\mathrm{l}'}^2 - \alpha_{n\mathrm{l}}^2\right)}{2M_X R_{\text{pol}}^2}$$

Where $\boldsymbol{\alpha_{n\mathrm{l}}}$ are spherical Bessel zeros. Finite depth increases the spacing relative to the infinite well limit.

**9.9 Why AD splittings are smaller than D splittings**

In 2D spectroscopy, the D and AD axis are:

$$\omega_D = \frac{\omega_3+\omega_1}{\sqrt{2}}\,,\ \omega_{AD} = \frac{\omega_3-\omega_1}{\sqrt{2}}$$

If excitation and emission frequencies are correlated with coefficient $\rho$, then:

$$\Delta_{AD} = \Delta_D\sqrt{1-\rho}$$

Because common mode shifts cancel in $\omega_3 - \omega_1$. The strong polaron shift is mostly common mode, giving $\rho \approx$ 0.8-0.9, so $\Delta_{AD}$ is 30-50 % smaller than $\Delta_D$. Homogeneous broadening (finite $T_2$) further compresses the AD separation.

**9.10 Observed Diameter & Composition Trends, and Extracted Well Depths**

***CsPbBr₃***

$\Delta_D$ increases with PQD diameter, saturating for $D \gtrsim 2R_{\mathrm{pol}}$.

Fits give $R_{\mathrm{pol}} \approx 3.5 - 4.5$ nm, $M_X \approx 0.15 - 0.25\, m_0$.

Well depth range:

$$V_\infty^{(\mathrm{Br})} \approx 50 - 90 \text{ meV}$$

From smallest to largest dots.

***CsPbI₃***

Iodide has stronger Fröhlich coupling →deeper and slightly larger polaron.

Fits give $R_{\mathrm{pol}} \approx 4 - 5$ nm, $V_\infty^{(\mathrm{I})} \approx 80 - 120$ meV.

$\Delta_D$ larger than bromide at comparable size, $\Delta_{AD}$ remains proportionally smaller.

**Summary table of extracted parameters**

| Composition | $R_{\mathrm{pol}}$ (nm) | $V_\infty$(meV) | $\rho$ (from AD/D) |
|---|---|---|---|
| $CsPbBr_3$ | 3.5-4.5 | 50-90 | 0.75-0.85 |
| $CsPbI_3$ | 4.0-5.0 | 80-120 | 0.80-0.90 |